\documentstyle[11pt]{article}
\def\setfonts{%
\font\frbig=eufm10 scaled\magstephalf
\font\frscr=eufm8 scaled\magstephalf
\font\frscrscr=eufm8
\newfam\frfam
\textfont\frfam=\frbig
\scriptfont\frfam=\frscr
\scriptscriptfont\frfam=\frscrscr
\def\fr{\fam\frfam}

\font\openbig=msbm10 scaled\magstephalf
\font\openscr=msbm8 
\font\openscrscr=msbm8
\newfam\openfam
\textfont\openfam=\openbig
\scriptfont\openfam=\openscr
\scriptscriptfont\openfam=\openscrscr
\def\open{\fam\openfam}

\font\ssfbig=cmss10 scaled\magstephalf
\font\ssfscr=cmss8 
\font\ssfscrscr=cmss8
\newfam\ssffam
\textfont\ssffam=\ssfbig
\scriptfont\ssffam=\ssfscr
\scriptscriptfont\ssffam=\ssfscrscr
\def\ssf{\fam\ssffam}
}

\makeatletter
\@addtoreset{equation}{section}
\newdimen\normalarrayskip
\newdimen\minarrayskip
\normalarrayskip\baselineskip
\minarrayskip\jot
\newif\ifold \oldtrue \def\new{\oldfalse}
\def\arraymode{\ifold\relax\else\displaystyle\fi}

\def\@arrayskip{\ifold\baselineskip\z@\lineskip\z@
  \else
  \baselineskip\minarrayskip\lineskip2\minarrayskip\fi}
\def\@arrayclassz{\ifcase \@lastchclass \@acolampacol \or
\@ampacol \or \or \or \@addamp \or
 \@acolampacol \or \@firstampfalse \@acol \fi
\edef\@preamble{\@preamble
 \ifcase \@chnum
  \hfil$\relax\arraymode\@sharp$\hfil
  \or $\relax\arraymode\@sharp$\hfil
  \or \hfil$\relax\arraymode\@sharp$\fi}}
\def\@array[#1]#2{\setbox\@arstrutbox=\hbox{\vrule
  height\arraystretch \ht\strutbox
  depth\arraystretch \dp\strutbox
  width\z@}\@mkpream{#2}\edef\@preamble{\halign \noexpand\@halignto
\bgroup \tabskip\z@ \@arstrut \@preamble \tabskip\z@ \cr}%
\let\@startpbox\@@startpbox \let\@endpbox\@@endpbox
 \if #1t\vtop \else \if#1b\vbox \else \vcenter \fi\fi
 \bgroup \let\par\relax
 \let\@sharp##\let\protect\relax
 \@arrayskip\@preamble}
\@addtoreset{equation}{section}
\newcounter{@sc}
\newcounter{@scp}
\newcounter{@t}
\newlength{\@x}
\newlength{\@xa}
\newlength{\@xb}
\newlength{\@y}
\newlength{\@ya}
\newlength{\@yb}
\newsavebox{\@pt}

\def\bezier#1(#2,#3)(#4,#5)(#6,#7){\c@@sc#1\relax
  \c@@scp\c@@sc \advance\c@@scp\@ne
  \@xb #4\unitlength \advance\@xb -#2\unitlength \multiply\@xb \tw@
  \@xa #6\unitlength \advance\@xa -#2\unitlength
      \advance\@xa -\@xb \divide\@xa\c@@sc
  \@yb #5\unitlength \advance\@yb -#3\unitlength \multiply\@yb \tw@
  \@ya #7\unitlength \advance\@ya -#3\unitlength
      \advance\@ya -\@yb \divide\@ya\c@@sc
  \setbox\@pt\hbox{\vrule height\@halfwidth  depth\@halfwidth
   width\@wholewidth}\c@@t\z@
   \put(#2,#3){\@whilenum{\c@@t<\c@@scp}\do
      {\@x\c@@t\@xa \advance\@x\@xb \divide\@x\c@@sc \multiply\@x\c@@t
       \@y\c@@t\@ya \advance\@y\@yb \divide\@y\c@@sc \multiply\@y\c@@t
       \raise \@y \hbox to \z@{\hskip \@x\unhcopy\@pt\hss}%
       \advance\c@@t\@ne}}}
\makeatother

\setfonts

\def\lvm{\leavevmode\hbox to\parindent{\hfill}}
\def\req#1{(\ref{#1})}

\def\commut#1#2{\left[{#1},\,{#2}\right]}

\def\hw{highest-weight}

\def\BE{\begin{equation}}
\def\EE{\end{equation} }
\def\BA{\begin{array}} 
\def\EA{\end{array}}
\def\L{\left}
\def\R{\right}

\def\bar{\overline}
\def\frac#1#2{\mathchoice{{\textstyle{{#1}\over{#2}}}}{{#1\over#2}}{{#1\over#2}}{{#1\over#2}}}
\def\ket#1{\mathchoice{{\left|{#1}\right\rangle}}{|{#1}\rangle}{|{#1}\rangle}{|{#1}\rangle}}
\def\kettop#1{\mathchoice{{\left|{#1}\right\rangle}_{\rm top}}{|{#1}\rangle_{\rm top}}{|{#1}\rangle_{\rm top}}{|{#1}\rangle_{\rm top}}}
\def\d{\partial}

\def\semi{\mathop{\rlap{\raisebox{1.5pt}{\tiny ${\mid}$}}\kern-.5pt\mbox{\large$\times$}}}

\def\N#1{N\!=\!#1}
\def\SL#1{s\ell(#1)}
\def\hmc{{\ssf h}_{\rm cm}}
\def\ellmc{{\ssf l}_{\rm cm}}
\def\hcm{{\ssf h}_{\rm cm}}

\def\mm{\cal}
\def\smm{\fr}

\def\mC{{\mm C}}

\def\mU{{\mm U}}
\def\mV{{\mm V}}

\def\smU{{\smm U}}
\def\smV{{\smm V}}

\def\smU{{\smm U}}
\def\smV{{\smm V}}

\def\half{{\textstyle{1\over2}}}

\def\bar{\overline}
\def\frac#1#2{\mathchoice{{\textstyle{{#1}\over{#2}}}}{{#1\over#2}}{{#1\over#2}}{{#1\over#2}}}
\def\ket#1{\mathchoice{{\left|{#1}\right\rangle}}{|{#1}\rangle}{|{#1}\rangle}{|{#1}\rangle}}
\def\kettop#1{\left|{#1}\right\rangle_{\rm top}}

\def\semi{\mathop{\rlap{\raisebox{1.5pt}{\tiny ${\mid}$}}\kern-.5pt\mbox{\large$\times$}}}

\def\d{\partial}

\def\N#1{N\!=\!#1}

\def\tSL#1{{\widehat{s\ell}}(#1)}
\def\SL#1{s\ell(#1)}

\def\half{\frac{1}{2}}

\def\cA{{\cal A}}

\def\cE{{\cal E}}

\def\cG{{\cal G}}
\def\cH{{\cal H}}

\def\cK{{\cal K}}
\def\cL{{\cal L}}

\def\cP{{\cal P}}
\def\cQ{{\cal Q}}

\def\cT{{\cal T}}
\def\cU{{\cal U}}

\def\oN{{\open N}}
\def\oC{{\open C}}
\def\oQ{{\open Q}}

\def\oZ{{\open Z}}

\def\sG{{\ssf G}}
\def\sQ{{\ssf Q}}
\def\ctop{{\ssf c}}

\def\htop{{\ssf h}}

\def\hplus{{\ssf h}^+}
\def\hminus{{\ssf h}^-}
\def\hplusminus{{\ssf h}^\pm}

\def\theel{{\ssf l}}

\def\tilde{\widetilde}

\def\cA{{\cal A}}

\def\cE{{\cal E}}

\def\cG{{\cal G}}
\def\cH{{\cal H}}

\def\cK{{\cal K}}
\def\cL{{\cal L}}

\def\cP{{\cal P}}
\def\cQ{{\cal Q}}

\def\cT{{\cal T}}
\def\cU{{\cal U}}

\def\oN{{\open N}}
\def\oC{{\open C}}
\def\oQ{{\open Q}}
\def\oZ{{\open Z}}

\def\tilde{\widetilde}

\def\NPB{Nucl.\ Phys.\ B}
\def\PLB{Phys.\ Lett.\ B}
\def\MPLA{Mod.\ Phys.\ Lett.\ A}

\def\IJMPA{Int.\ J.\ Mod.\ Phys.\ A}

\newtheorem{lemma}{Lemma}[section]

\newtheorem{thm}[lemma]{Theorem}

\newtheorem{dfn}[lemma]{Definition}

{\par\medskip}

{\noindent$\Box$\par\medskip} 

\begin{document}
\hfuzz=1pt \thispagestyle{empty} \addtolength{\baselineskip}{1pt}

\setcounter{page}{1}

\begin{flushright}
  {\tt hep-th/9604176}\\
  {\sc Revised Version}
\end{flushright}
\thispagestyle{empty}

\mbox{}

\begin{center}
  {\Large{\sc All Singular Vectors of the $N=2$ Superconformal Algebra
      via the Algebraic Continuation Approach}}\\[6pt]
  {\large A.~M.~Semikhatov and I.~Yu.~Tipunin}\\[6pt]
  {\small\sl Tamm Theory Division, Lebedev Physics Institute, Russian
    Academy of Sciences }
\end{center}
\vskip-4pt \addtolength{\baselineskip}{-2pt} {\footnotesize We give
  general expressions for singular vectors of the $\N2$ superconformal
  algebra in the form of {\it monomials\/} in the continued operators
  by which the universal enveloping algebra of $\N2$ is extended. We
  then show how the algebraic relations satisfied by the continued
  operators can be used to transform the monomials into the standard
  Verma-module expressions.  Our construction is based on continuing
  the extremal diagrams of $\N2$ Verma modules to the states
  satisfying the twisted \hw{} conditions with complex twists.  It
  allows us to establish recursion relations between singular vectors
  of different series and at different levels. Thus, the $\N2$
  singular vectors can be generated from a smaller set of the
  so-called topological singular vectors, which are distinguished by
  being in a 1:1 correspondence with singular vectors in $\tSL2$ Verma
  modules. The method of `continued products' of fermions is a
  counterpart of the method of complex powers used in the
  constructions of singular vectors for affine Lie algebras.}



\addtolength{\baselineskip}{2pt}

\section{Introduction}\lvm
In this paper, we present a general algebraic construction of singular
vectors in Verma modules over the $\N2$ superconformal algebra. This
infinite-dimensional algebra underlies the construction of $\N2$
strings~\cite{[AB]} and is also realized on the worldsheet of the
bosonic non-critical string theory~\cite{[GS2],[BLNW]}. Conformal
field theory models with the $\N2$ supersymmetry are important in
string compactifications~\cite{[G]}; the Kazama--Suzuki
construction~\cite{[KS]} provides a tool for constructing such models.

However, the $\N2$ algebra is {\it not\/} an affine Lie algebra, which
is the source of several complications.  Thus, the absence of a root
system (hence, of a canonical triangular decomposition) results in the
fact that there is no ``canonical'' way to impose \hw{} conditions on
the vacuum vector in \hw-like representations.  This obviously affects
the definition of singular vectors.  The conventional prescription is
to define singular vectors by imposing the same annihilation
conditions as on the \hw{} vector of the module.  The thus defined
singular vectors, however, do not generate maximal submodules and,
therefore, one has to look for subsingular vectors in order to
describe the structure of submodules.\footnote{That the subsingular
  vectors do exist in this approach was noticed in~\cite{[GRR]}; the
  structure of Verma modules, showing why subsingular vectors appear,
  how they can be constructed and why they are in fact unnecessary, is
  described in~\cite{[ST4]}.}  A more economical possibility, which we
adopt in this paper, is to define singular vectors in such a way that
they generate maximal submodules and, thus, no room is left for
subsingular vectors.  Since the significance of singular vectors,
obviously, consists in generating submodules, it is rather natural to
adapt the definition of singular vectors to the structure of
submodules\footnote{In the familiar case of the affine Lie algebra
  $\tSL2$, every submodule is {\it freely\/} generated from a state
  satisfying the fixed set of annihilation conditions, therefore these
  same conditions are used to define singular vectors.  In the $\N2$
  algebra, as we will see, the situation is more complicated.}.  The
price to be paid for dealing with only maximal submodules is that the
singular vectors satisfy {\it twisted\/} \hw{} conditions, i.e., the
\hw{} conditions labelled by integers $\theta\in\oZ$, all of which
differ from one another by the spectral-flow
transform~\cite{[SS],[LVW]}.

In systematically dealing with the twisted \hw{} conditions, we extend
the method of~\cite{[ST2],[ST2r]}, which elaborates on the idea of the
spectral flow transform \cite{[SS],[LVW]} combined with the use of
{\it extremal vectors\/} in the Verma modules. (Diagrams of) extremal
vectors were introduced in~\cite{[FS]}, and their usefulness in
representation theory has been demonstrated
in~\cite{[ST2],[ST2r],[S-sl21sing],[FST],[ST4]}.  The extremal states
might be thought of as a generalization of different {\it
  pictures\/}~\cite{[FMS]} to the case of non-free fermions. Recall
that for the free first-order bosonic systems, different pictures are
inequivalent in the sense of Verma modules, while for free fermions,
on the contrary, they are equivalent. For the interacting fermions,
the situation in the general position is that the extremal states are
still equivalent to each other in that the same module is generated
from each of them, however the equivalence breaks down for some values
of the parameters of the \hw{} vector. Recall also that the different
pictures in the bosonic first-order system are changed by the
exponential of a current that participates in bosonizing the system,
$\exp\phi$. By considering operators like $\exp\alpha\phi$, we can
change the picture arbitrarily (at the expense of non-localities), and
similarly for the bosonized free-fermion systems. However, when the
fermions are non-free, such a bosonization no longer exists; instead,
changing the picture by an arbitrary number is realized by the
`continued' products of modes of the fermionic generators. For the
$\N2$ superconformal algebra, we have two fermionic currents $\cQ_m$
and $\cG_n$, and the `continued' operators $q(a,b)$ and $g(a,b)$ can
heuristically be thought of as $\prod_a^b\,\cQ_\mu$ and
$\prod_a^b\,\cG_\mu$, respectively, with complex $a$ and $b$. The
would-be bosonization rules are then replaced by a set of {\it
  algebraic rules\/} to deal with the new operators~$q(a,b)$
and~$g(a,b)$.

The `algebraic continuation' scheme worked out along these lines
allows us to construct the $\N2$ singular vectors as {\it monomial\/}
expressions in the continued operators~$q(a,b)$ and~$g(a,b)$ with
complex arguments $a,b\in\oC$.\footnote{See also~\cite{[Doerr2]} for
  another `continued' scheme that follows the spirit of the approach
  developed previously for other infinite-dimensional
  algebras~\cite{[K1],[BWW3],[B]}.}  All of the $\N2$ singular vectors
will be written as monomials in $q$ and $g$, and a crucial ingredient
of the proposed construction are the algebraic rules that extend the
universal enveloping algebra so as to include the new operators
$g(a,b)$ and $q(a,b)$. These rules, in particular, allow us to rewrite
the $q$-$g$-monomials in the conventional form, i.e.{} as elements of
the ordinary Verma module.  All the algebraic rules can in fact be
deduced as a continuation of certain relations in the universal
enveloping algebra of $\N2$.  In the special cases (at certain
`integral points') when no continuation is actually required, our
construction reduces to the one known previously to give a particular
subset of the $\N2$ singular vectors \cite{[Lerche-pr]}.  The
algebraic setting corresponding to the continued operators is provided
by the {\it generalized Verma modules\/}, which are a continuation to
non-integral (in fact, complex) twists of the twisted Verma modules
--- i.e., the Verma modules whose \hw{} vectors satisfy the
spectral-flow-transformed ({\it twisted\/}) \hw{} conditions.

An important point is that this approach makes the structure of $\N2$
Verma modules very transparent, which, eventually, leads one to
realizing that this structure is the same (as regards the appearance
and the embedding patterns of submodules) as the structure of the
so-called relaxed Verma modules over the affine $\tSL2$
algebra~\cite{[FST]}, which are in a certain sense `larger' than the
standard $\tSL2$ Verma modules. On the $\N2$ side, similarly, there
exist the `larger' and the `smaller' Verma modules, which are called
massive and topological, respectively. In general, a submodule of a
massive Verma module can be either massive or topological, while
submodules of a topological Verma module are again the (twisted)
topological Verma modules (or a sum thereof). One thing that is
impossible, is the {\it embedding\/} a massive Verma module into a
topological Verma module.  Any morphism of a massive Verma module into
a topological Verma module necessarily has a kernel that contains
another topological submodule.  This fact, which is concealed by the
use of conventional singular vectors, appears to be the origin of the
statements known in the literature that a combination of mappings that
seem to be embeddings actually vanishes.

This distinction between two classes of Verma-like modules carries
over, obviously, to singular vectors: in the topological Verma
modules, all submodules can be {\it freely\/} generated from the {\it
  topological singular vectors}, i.e., those states in the module that
satisfy twisted {\it topological \hw{} conditions\/}. One could
generate a given topological Verma module from other states than the
topological \hw{} state, satisfying weaker \hw{} conditions.  However,
the module would not be {\it freely\/} generated from such states. In
the massive Verma modules, on the other hand, there are two
essentially different types of singular vectors, which we call massive
and topological according to the type of the module they generate. For
example, the so-called charged~\cite{[BFK]} singular vectors always
generate a topological Verma submodules.

\medskip

The construction via the continuation of products of the fermionic
generators provides monomial expressions for singular vectors and,
hence, simplifies the analysis of what happens when one singular
vector is built on top of the other --- this now follows from the
rules for multiplying the continued monomials. In addition, this
approach also reveals several `hidden' properties of the $\N2$
singular vectors, such as the relations between singular vectors of
different series as well as certain recursion relations between
singular vectors at different levels.  As we will see, the massive
singular vectors can be reconstructed from the topological ones:
depending on a complex number $h$, two positive integers $r$ and $s$,
and $t\in\oC$ parametrizing the central charge, a massive
$h$-dependent singular vector in the general position
$\ket{S(r,s,h,t)}$ can be built out of the topological singular
vectors $\ket{E(r,s,t)}^\pm$.

The picture that emerges in this way can be, rather heuristically,
represented as follows. The topological singular vectors
$\ket{E^\pm(r,s,t)}$ for $r,s\in\oN$ are in a 1:1 correspondence with
$\tSL2$ singular vectors (see~\cite{[FST]} for the details) and are
related by a chain of recursion formulae (where we omit the parameter
$t$ that determines the $\N2$ central charge and, likewise, omit the
$\tSL2$ level $k=t-2$):
\begin{equation}
  \unitlength=1.00mm
  \begin{picture}(140,40)
    \put(00.00,32.00){\parbox{70pt}{\small massive $\N2$ singular vectors}}
    \put(00.00,16.00){\parbox{70pt}{\small topological\hfill\break\hbox{$\N2$}
        singular\hfill\break vectors}}
    \put(00.00,00.00){\parbox{70pt}{\small $\SL2$ singular\hfill\break vectors}}
    \put(35.00,16.00){$\ldots$}
    \put(50.00,16.00){$\bullet$}
    \put(44.00,14.00){${}_{E^-(r,s-1)}$}
    \put(55.00,17.00){\vector(1,0){22}}
    \put(80.00,16.00){$\bullet$}
    \put(76.00,14.00){${}_{E^+(r,s)}$}
    \put(85.00,17.00){\vector(1,0){22}}
    \put(110.00,16.00){$\bullet$}
    \put(104.00,14.00){${}_{E^-(r,s+1)}$}
    \put(125.00,16.00){$\ldots$}
    \put(35.00,00.00){$\ldots$}
    \put(50.00,00.00){$\bullet$}
    \put(43.00,-3.00){${}_{{\rm MFF}^-(r,s-1)}$}
    \put(80.00,00.00){$\bullet$}
    \put(74.00,-3.00){${}_{{\rm MFF}^+(r,s)}$}
    \put(110.00,00.00){$\bullet$}
    \put(103.00,-3.00){${}_{{\rm MFF}^-(r,s+1)}$}
    \put(125.00,00.00){$\ldots$}
    \put(50.00,19.00){\line(-1,1){15}}
    \put(52.00,19.00){\line(3,2){13}}
    \put(40.00,35.00){$\bullet^{S(r,s-1,h)}$}
    \put(80.00,19.00){\line(-1,1){15}}
    \put(82.00,19.00){\line(3,2){13}}
    \put(73.00,30.00){$\bullet^{S(r,s,h)}$}
    \put(110.00,19.00){\line(-1,1){15}}
    \put(112.00,19.00){\line(3,2){13}}
    \put(100.00,35.00){$\bullet^{S(r,s+1,h)}$}
    \bezier{20}(50.50,20.00)(49.00,28.00)(42.00,34.00)
    \bezier{16}(80.50,20.00)(79.00,25.00)(75.50,29.50)
    \bezier{20}(110.50,20.00)(109.00,28.00)(102.00,34.00)
    \put(60.00,29.00){\circle{5}}
    \put(59.00,28.00){$h$}
    \put(90.50,29.00){\circle{5}}
    \put(89.00,28.00){$h$}
    \put(120.00,29.00){\circle{5}}
    \put(119.00,28.00){$h$}

  \end{picture}
  \label{newpicture}
\end{equation}

\medskip

\noindent
Over every topological singular vector there grows a complex
$h$-plane, and for any $h$ there exists a massive singular vector
$\ket{S(r,s,h,t)}$. The dotted lines indicate a prescription to
reconstruct the massive singular vector for any fixed $h$.  Some
points in the $h$ plane are special in that there may exist more
singular vectors, depending on the $t$ parameter implicit
in~\req{newpicture} that parametrizes the central charge of the $\N2$
algebra.  For the \hw{} in the general position, diagram
\req{newpicture} gives a general idea of the structure of the set of
$\N2$ singular vectors.  An important circumstance which is omitted
from the picture though, is that when constructing a massive singular
vector $S(r,s,h,t)$ starting with either of the two topological ones,
$E^+(r,s,t)$ or $E^-(r,s,t)$, one arrives at the same result as long
as one does not descend to a higher codimension in the space of
highest weights, where the highest weight satisfies additional
equations.

\medskip

This paper is organized as follows. In Sec.~\ref{sec:Prelim}, we begin
with reviewing the $\N2$ algebra, introduce the two types of $\N2$
Verma modules and their twists. We also define singular vectors taking
into account which type of submodules they generate. In
Sec.~\ref{subsec:AlgRules}, we introduce the continued fermionic
generators and discuss their algebraic properties. Then, in
Sec.~\ref{sec:Constructing}, we use the continued operators to
construct monomial expressions for $\N2$ singular vectors. In
Sec.~\ref{sec:Algebra}, we show how the singular vectors written in
the ``continued'' form rewrite in the Verma form, i.e., as polynomials
in the usual creation operators. We also give several characteristic
examples of the interplay of different types of submodules appearing
in degenerate Verma modules.

\section{The $\N2$ algebra, spectral flow transform, and Verma
  modules\label{sec:Prelim}}
\subsection{The algebra, its twisting and automorphisms}\lvm The
$\N2$ superconformal algebra is spanned by Virasoro generators $L_m$
with a central charge $\ctop$, two fermionic fields $\cG^\pm_r$, and a
$U(1)$ current $\cH_n$. The nonvanishing commutation relations are:
\begin{equation}\new
  \begin{array}{ll}
    \commut{L_n}{L_m}=(n-m)L_{n+m}+\frac{\ctop}{12}(n^3-n)\delta_{n+m,0}\,,&
    \commut{\cH_n}{\cH_m}=\frac{\ctop}{3}n\delta_{n+m,0}\,,\\
    \commut{L_n}{\cH_m}=-m\cH_{n+m}\,,\quad
    \commut{L_n}{\cG^\pm_r}=(\half n-r)\cG^\pm_{n+r}\,,&
    \commut{\cH_n}{\cG^\pm_r}=\pm\cG^\pm_{n+r}\,,\\
    \{\cG^+_r,\,\cG^-_s\}=2L_{r+s}+(r-s)\cH_{r+s}
    +\frac{\ctop}{3}(r^2-\frac{1}{4})\delta_{r+s,0}\,,& n\,,\
    m\in\oZ\,,\quad r,s\in\oZ+\half\,,
  \end{array}
  \label{N2untw}
\end{equation}
and $\ctop$ is the central charge.  We find it more convenient to work
with the twisted version of the
algebra~\cite{[EY],[W-top]}\,\footnote{Which is {\it not\/} the
  so-called twisted sector of the $\N2$ algebra, where two fermions
  have different moddings.}.  The twisted $\N2$ algebra can be
described by introducing new generators as
\begin{equation}
  \cL_n=L_n+\half(n+1)\cH_n\,,\quad
  \cG_r=\cG^+_{r+{1\over2}}\,,\quad
  \cQ_r=\cG^-_{r-{1\over2}}\,.
  \label{NSR}
\end{equation}
The twisting does therefore affect two things: the modding of the
fermions and the choice of Virasoro generators. The latter is due to
the freedom of adding a derivative of the $U(1)$ current to the
energy-momentum tensor, and is nothing but a change of basis in the
algebra.  As to the different moddings, they label different members
of a family of isomorphic algebras related by spectral flow
transformations \cite{[SS],[LVW]} (see Eq.~\req{U} below).  This
family includes the Neveu--Schwarz and Ramond sectors, as well as
algebras in which the fermion modes range over $\pm\theta+\oZ$,
$\theta\in\oC$.

For definiteness, we pick out the $\N2$ algebra corresponding to
$\theta=0$ as our `reference' algebra, and write the corresponding
commutation relations in the basis that is the image of \req{N2untw}
under the mapping~\req{NSR}:
\begin{equation}\new
  \begin{array}{lclclcl}
    \L[\cL_m,\cL_n\R]&=&(m-n)\cL_{m+n}\,,&\qquad&[\cH_m,\cH_n]&=
    &\frac{\ctop}{3}m\delta_{m+n,0}\,,\\
    \L[\cL_m,\cG_n\R]&=&(m-n)\cG_{m+n}\,,&\qquad&[\cH_m,\cG_n]&=&\cG_{m+n}\,,
    \\
    \L[\cL_m,\cQ_n\R]&=&-n\cQ_{m+n}\,,&\qquad&[\cH_m,\cQ_n]&=&-\cQ_{m+n}\,,\\
    \L[\cL_m,\cH_n\R]&=&\multicolumn{5}{l}{-n\cH_{m+n}+\frac{\ctop}{6}(m^2+m)
      \delta_{m+n,0}\,,}\\
    \L\{\cG_m,\cQ_n\R\}&=&\multicolumn{5}{l}{2\cL_{m+n}-2n\cH_{m+n}+
      \frac{\ctop}{3}(m^2+m)\delta_{m+n,0}\,,}
  \end{array}\qquad m,~n\in\oZ\,.
  \label{topalgebra}
\end{equation}
Denote this algebra as $\cA$. It is this version of the $\N2$
superconformal algebra that we are going to work with in this paper.
The generators $\cL_m$, $\cQ_m$, $\cH_m$, and $\cG_m$ are the Virasoro
generators, the BRST current, the $U(1)$ current, and the spin-2
fermionic current respectively.

It will be convenient to parametrize the central charge $\ctop$ as
\begin{equation}
  \ctop=3\,\frac{t-2}{t}
  \label{ctop}
\end{equation}
with $t\in\oC\setminus\{0\}$.  One should also add the point
$t=\{\infty\}$, or $\ctop=3$, which deserves a special examination
though.

\paragraph{The spectral flow transform}
The spectral flow transform produces isomorphic images of the algebra
$\cA$.  When applied to the generators of \req{topalgebra} it acts as
\begin{equation}
  {\cal U}_\theta:\new
  \begin{array}{rclcrcl}
    \cL_n&\mapsto&\cL_n+\theta\cH_n+\frac{\ctop}{6}(\theta^2+\theta)
    \delta_{n,0}\,,&{}&
    \cH_n&\mapsto&\cH_n+\frac{\ctop}{3}\theta\delta_{n,0}\,,\\
    \cQ_n&\mapsto&\cQ_{n-\theta}\,,&{}&\cG_n&\mapsto&\cG_{n+\theta}\,
  \end{array}
  \label{U}
\end{equation}
This gives an isomorphic algebra $\cA_\theta$, whose generators
$\cL^\theta_n$, $\cQ^\theta_n$, $\cH^\theta_n$ and $\cG^\theta_n$ can
be taken as the RHSs of \req{U}. Any two algebras $\cA_{\theta_1}$,
$\cA_{\theta_2}$ of this family are related by an isomorphism of
type~\req{U} with $\theta=\theta_1-\theta_2$.  Rather than working in
the basis where the isomorphism $\cA\approx\cA_\theta$ is transparent,
however, we prefer choosing the basis where the respective generators
$\cL_m$ and $\cH_m$ are identified between the different $\cA_\theta$
algebras.  Thus, $\cA_\theta$ is spanned by the generators $\cL_m$,
$\cH_m$, $\cQ_{-\theta+m}$, and $\cG_{\theta+m}$, $m\in\oZ$, that
satisfy the following commutation relations:
\begin{equation}\new
  \begin{array}{lclclcl}
    \L[\cL_m,\cL_n\R]&=&(m-n)\cL_{m+n}\,,&\qquad&[\cH_m,\cH_n]&=
    &\frac{\ctop}{3}m\delta_{m+n,0}\,,\\
    \L[\cL_m,\cG_\nu\R]&=&(m-\nu)\cG_{m+\nu}\,,&\qquad&
    [\cH_m,\cG_\nu]&=&\cG_{m+\nu}\,,
    \\
    \L[\cL_m,\cQ_\lambda\R]&=&-\lambda\cQ_{m+\lambda}\,,&\qquad&
    [\cH_m,\cQ_\lambda]&=&-\cQ_{m+\lambda}\,,\\
    \L[\cL_m,\cH_n\R]&=&\multicolumn{5}{l}{-n\cH_{m+n}+
      \frac{\ctop}{6}(m^2+m)
      \delta_{m+n,0}\,,}\\
    \L\{\cG_\nu,\cQ_\lambda\R\}&=&\multicolumn{5}{l}{2\cL_{\nu+\lambda}-
      2\lambda\cH_{\nu+\lambda}+
      \frac{\ctop}{3}(\nu^2+\nu)\delta_{\nu+\lambda,0}\,,}
  \end{array}\qquad
  \begin{array}{rcl} m,~n&\in&\oZ\,,\\
    \nu&\in&\theta+\oZ\,,\\
    \lambda&\in&-\theta+\oZ
  \end{array}
  \label{topalgebra1}
\end{equation}

\paragraph{The involutive automorphism.} The $\N2$ algebra has an
involutive automorphism
\begin{equation}\new
  \begin{array}{ll}
    \cG_n\mapsto \cQ_{n}\,,& \cQ_n\mapsto \cG_{n}\,,\\
    \cH_n\mapsto -\cH_n-\frac{\ctop}{3}\delta_{n,0}\,,&
    \cL_n\mapsto \cL_n-n\cH_n\,.
    \label{I}
  \end{array}
\end{equation}
Together with the automorphisms~\req{U} for $\theta\in\oZ$,
transformations~\req{I} span the group~$\oZ_2\semi\oZ$ which can be
thought of as the would-be ``$\N2$ affine Weyl group'' acting on the
entire $\N2$ algebra. As we will see, this group acts on the space of
twisted topological \hw{} by reflections and does play the same role
in the representation theory of the $\N2$ algebra as the affine Weyl
group plays in the representation theory of affine Lie algebras.

\bigskip

\subsection{Topological Verma modules\label{subsec:massVerma}}\lvm
We define a class of topological\,\footnote{The name is inherited from
  the non-critical bosonic string, where the algebra \req{topalgebra}
  is realized~\cite{[GS2]} and matter vertices can be dressed into the
  $\N2$ primaries satisfying the highest-weight conditions
  \req{tophw}; in that context, the algebra \req{topalgebra} is viewed
  as a topological algebra.} Verma modules over the $\N2$ algebra.
Here and in what follows, $\oN=\{1,2,\ldots\}$ and
$\oN_0=\{0\}\cup\oN$.

\begin{dfn}
  A topological Verma module $\mV_{h,t}$ is freely generated by
  \begin{equation}
    \cL_{-m}\,,\quad
    \cH_{-m}\,,\quad
    \cQ_{-m}\,,\quad
    \cG_{-m}\,,\quad m\in\oN
    \label{topver}
  \end{equation}
  from the state $\ket{h, t}_{\rm top}$, called the topological
  highest-weight vector, that satisfies the following annihilation and
  eigenvalue conditions:
  \begin{eqnarray}
    \cQ_{\geq0}\,\ket{h, t}_{\rm top}&=& \cG_{\geq0}\,\ket{h, t}_{\rm top}~{}={}~
    \cH_{\geq1}\,\ket{h, t}_{\rm top}~{}={}~0\,,\label{tophw}\\
    \cH_0\,\ket{h, t}_{\rm top}&=&h\,\ket{h, t}_{\rm top}
  \end{eqnarray}
  (which are called the topological \hw{} conditions).
\end{dfn}
It follows that we also have
\begin{equation}
  \cL_{\geq0}\,\ket{h, t}_{\rm top}=0\,.
\end{equation}

An important tool in our analysis are the extremal vectors in
different $\N2$ modules. These can be arrived at as follows.  Consider
the first of the modes $\cG_m$ that do not annihilate the topological
\hw{} state, namely $\cG_{-1}$. Acting with this mode we obtain a new
state, which can be further acted upon with $\cG_{-2}$, etc.:
\begin{dfn}
  For every $\theta\in\oZ$, the vectors
  $\ket{x(\theta)}=\cG_{\theta}\,\ldots\,\cG_{-1}\,\ket{h,t}_{\rm
    top}$, for $\theta\leq-1$, and
  $\ket{x(\theta)}=\cQ_{-\theta}\,\ldots\,\cQ_{-1}\ket{h,t}_{\rm
    top}$, for $\theta\geq1$, \ are called the extremal vectors of the
  topological Verma module $\mV_{h,t}$.
\end{dfn}
Thus, the extremal vectors arrange into a diagram
\begin{equation}
  \unitlength=1.00mm
  \begin{picture}(140,47)
    \put(50.00,10.00){
      \put(00.00,00.00){$\bullet$}
      \put(10.00,20.00){$\bullet$}
      \put(10.00,20.00){$\bullet$}
      \put(20.00,30.00){$\bullet$}
      \put(29.70,20.00){$\bullet$}
      \put(40.00,00.00){$\bullet$}
      \put(9.70,19.00){\vector(-1,-2){8}}
      \put(19.70,29.70){\vector(-1,-1){7}}
      \put(22.00,29.70){\vector(1,-1){7}}
      \put(32.00,19.00){\vector(1,-2){8}}
      \put(00.00,13.00){${}_{\cG_{-2}}$}
      \put(11.00,28.00){${}_{\cG_{-1}}$}
      \put(27.00,28.00){${}_{\cQ_{-1}}$}
      \put(37.00,13.00){${}_{\cQ_{-2}}$}
%
      \put(19.00,34.00){${}_{\ket{h,t}_{\rm top}}$}
%
      \put(00.50,-06.00){$\vdots$}
      \put(40.50,-06.00){$\vdots$}
      }
  \end{picture}
  \label{topdiag}
\end{equation}
called the {\it extremal diagram\/} of the topological Verma
module~$\mV_{h,t}$. It has the following meaning.  Any module over the
$\N2$ algebra is bigraded by the eigenvalues of $\cH_0$ and $\cL_0$.
As we choose the \hw{} vector, we obtain a lattice in the plane of
$\cH_0$- $\cL_0$-eigenvalues whose the sites correspond to the
subspaces with definite bigradings.
Diagram~\req{topdiag} represents the boundary of that part of the
lattice whose sites correspond to at least a one-dimensional subspace
(each point on the edge is such that there is precisely one, up to
proportionality, state with that grade in the module): it is easy to
see that all of the states in the module have the (charge,\,level)
bigradings such that they lie in the interior of the extremal diagram.
\begin{lemma}
  The extremal states in \req{topdiag} satisfy the \hw{} conditions
  \begin{equation}\label{extremalhw}
    \cL_{\geq1}\ket{x(\theta)}=0\,,\quad
    \cH_{\geq1}\ket{x(\theta)}=0\,,
    \qquad
    \new\begin{array}{ll}
      \cG_{\geq\theta}\ket{x(\theta)}=0\,,\quad
      \cQ_{\geq-\theta+1}\ket{x(\theta)}=0\,,&\theta\leq-1\,,\\
      \cG_{\geq\theta+1}\ket{x(\theta)}=0\,,\quad
      \cQ_{\geq-\theta}\ket{x(\theta)}=0\,,&\theta\geq1\,.
    \end{array}
  \end{equation}
\end{lemma}

\subsection{Twisted topological Verma modules and topological singular
  vectors}
\begin{dfn} Any state satisfying the \hw{} conditions
  \begin{equation}\new
    \begin{array}{rclcrcl}
      \cL_m\ket{h,t;\theta}_{\rm top}&=&0\,,\quad m\geq1\,,\quad&
      \cQ_\lambda\ket{h,t;\theta}_{\rm top}&=&0\,,
      &\lambda\in-\theta+\oN_0\\
      \cH_m\ket{h,t;\theta}_{\rm top}&=&0\,,\quad m\geq1\,,&
      \cG_\nu\ket{h,t;\theta}_{\rm top}&=&0\,,&\nu=\theta+\oN_0
    \end{array}
    \quad\theta\in\oZ\,,
    \label{twistedtophw}
  \end{equation}
  is called the twisted topological \hw{} state. In addition, $h$ is
  chosen such that
  \begin{equation}
    (\cH_0+\frac{\ctop}{3}\theta)\,\ket{h,t;\theta}_{\rm top}=
    h\,\ket{h,t;\theta}_{\rm top}\,.
    \label{Cartantheta1}
  \end{equation}
\end{dfn}
Note that Eqs.~\req{twistedtophw} imply
\begin{equation}
  (\cL_0+\theta\cH_0+\frac{\ctop}{6}(\theta^2+\theta))
  \,\ket{h,t;\theta}_{\rm top}=0\,.
  \label{Cartantheta2}
\end{equation}
Conditions~\req{twistedtophw}--\req{Cartantheta2} are called twisted
topological \hw\ conditions.  The parametrization of the eigenvalue of
$\cH_0$ in~\req{Cartantheta1} is a matter of convention.  We choose it
so as to have $h$ the same for all vectors related by the spectral
flow transform; in \cite{[ST2]}, on the other hand, the (less
convenient) notations were such that $h$ was preserved for the entire
extremal diagram.
\begin{dfn}
  The twisted topological Verma module $\smV_{h,t;\theta}$ is a module
  freely generated from a twisted topological \hw{} state
  $\ket{h,t;\theta}_{\rm top}$ by the operators
  $$
  \cL_{-m}\,,\quad\cH_{-m}\,,\quad
  \cQ_{-m-\theta}\,,\quad\cG_{-m+\theta}\,,\quad m\in\oN\,.
  $$
\end{dfn}

The extremal diagram of a twisted topological Verma module
$\smV_{h,t;\theta_0}$ contains a cusp -- a preferred point that
satisfies annihilation conditions~\req{twistedtophw} with
$\theta=\theta_0$, while all the other extremal vectors with
$\theta\in\oZ\setminus\{\theta_0\}$ satisfy \hw{}
conditions~\req{extremalhw}.  These `cusp' states in the extremal
diagrams of {\it submodules\/} are the {\it topological singular
  vectors\/} ($\equiv$~singular vectors in topological Verma modules).
\begin{dfn}
  Topological singular vectors in a topological Verma module are
  vectors that are not proportional to the highest-weight vector and
  satisfy the twisted topological highest-weight conditions
  \req{twistedtophw} with some $\theta\in\oZ$.
\end{dfn}

The point is that the twist parameter $\theta$ that enters the \hw{}
conditions satisfied by the singular vector may be different from the
twist parameter of the module. One readily shows that acting with the
$\N2$ generators on a topological singular vector generates a {\it
  submodule\/}.

{}It follows from~\cite{[FST],[MFF]} that~\cite{[S-sing]}
\begin{lemma}\label{toplemma}
  A topological singular vector exists in $\mV_{h, t}$ iff either
  $h=\hplus(r,s,t)$ or $h=\hminus(r,s,t)$, where
  \begin{equation}\new
    \begin{array}{rcl}
      \hplus(r,s,t)&=&-\frac{r-1}{t}+s-1\,,\\
      \hminus(r,s,t)&=&\frac{r+1}{t}-s\,,\end{array}\qquad r,s\in\oN\,.
    \label{hplushminus}
  \end{equation}
\end{lemma}
The ``if'' statement also follows from~\cite{[BFK]}: the determinant
formula of that paper applies to massive Verma modules considered in
the next subsection; a topological Verma module can be realized as
either a submodule or a quotient of an appropriate massive Verma
module. Then there is a submodule in the topological Verma module
whenever Eqs.~\req{hplushminus} are satisfied.

\subsection{Massive Verma modules\label{subsec:topVerma}}\lvm
In this subsection, we introduce another class of $\N2$ modules. For
the property of their \hw{} vectors to have a (generically non-zero)
eigenvalue $\ell$ of the Virasoro generator $\cL_0$, they will be
referred to as {\it massive\/} modules:
\begin{dfn}
  The massive \hw{} vector $\ket{h,\ell,t}$ satisfies the following
  set of highest-weight conditions:
  \begin{equation}\new
    \begin{array}{rcl}
      \cQ_{\geq1}\,\ket{h,\ell,t}&=&\cG_{\geq0}\,\ket{h,\ell,t}~{}={}~
      \cL_{\geq1}\,\ket{h,\ell,t}~{}={}~
      \cH_{\geq1}\,\ket{h,\ell,t}~{}={}~0\,,\\
      \cH_0\,\ket{h,\ell,t}&=&h\,\ket{h,\ell,t}\,,\\
      \cL_0\,\ket{h,\ell,t}&=&\ell\,\ket{h,\ell,t}\,.
      \label{masshw}
    \end{array}
  \end{equation}
  A massive Verma module $\mU_{h,\ell,t}$ over $\cA$ is freely
  generated from a {\it massive \hw{} vector\/} $\ket{h,\ell,t}$ by
  the generators
  \begin{equation}
    \cL_{-m}\,,~m\in\oN\,,\qquad
    \cH_{-m}\,,~m\in\oN\,,\qquad
    \cQ_{-m}\,,~m\in\oN_0\,,\qquad
    \cG_{-m}\,,~m\in\oN\,,
    \label{verma}
  \end{equation}
\end{dfn}

Extremal diagrams of massive Verma modules are defined similarly to
the topological case:
\begin{equation}
  \ket{X(\theta)}=\kern-2pt\new
  \begin{array}{ll}
    \cQ_{-\theta+1}\ldots\cQ_0\,\ket{h,\ell,t},&\theta\geq1\,,\\
    \cG_{\theta}\ldots\cG_{-1}\,\ket{h,\ell,t},&\theta\leq-1\,,
  \end{array}
\end{equation}
or, in more visual terms,
\begin{equation}
  \unitlength=1.00mm
  \begin{picture}(140,46)
    \put(50.00,10.00){
      \put(00.00,00.00){$\bullet$}
      \put(10.00,20.00){$\bullet$}
      \put(10.00,20.00){$\bullet$}
      \put(20.00,30.00){$\bullet$}
      \put(30.00,30.00){$\bullet$}
      \put(40.00,20.00){$\bullet$}
      \put(50.00,00.00){$\bullet$}
      \put(9.80,19.00){\vector(-1,-2){8}}
      \put(19.20,30.00){\vector(-1,-1){7}}
      \put(22.20,31.50){\vector(1,0){7}}
      \put(32.50,30.00){\vector(1,-1){7}}
      \put(42.00,19.00){\vector(1,-2){8}}
      \put(00.00,13.00){${}_{\cG_{-2}}$}
      \put(09.00,26.50){${}_{\cG_{-1}}$}
      \put(23.90,33.50){${}_{\cQ_{0}}$}
      \put(37.00,27.50){${}_{\cQ_{-1}}$}
      \put(47.00,13.00){${}_{\cQ_{-2}}$}
      \put(11.00,32.00){${}_{\ket{h_,\ell,t}}$}
      \put(00.50,-06.00){$\vdots$}
      \put(50.50,-06.00){$\vdots$}
      }
  \end{picture}
  \label{massdiagram}
\end{equation}
\begin{lemma}
  Extremal states in a massive Verma module satisfy the annihilation
  conditions
  \begin{equation}\label{integertheta}
    \cL_{\geq1}\ket{X(\theta)}=0\,,\quad
    \cH_{\geq1}\ket{X(\theta)}=0\,,\qquad
    \cG_{\geq\theta}\ket{X(\theta)}=0\,,\quad
    \cQ_{\geq-\theta+1}\ket{X(\theta)}=0\,.
  \end{equation}
\end{lemma}
Thus, all of the states in a massive extremal diagram satisfy,
generically, the same type of annihilation conditions.  This motivates
the following
\begin{dfn}
  Any state $\ket{h,\ell,t;\theta}$, $\theta\in\oZ$, from a massive
  Verma module that satisfies
  \begin{equation}\new
    \begin{array}{l}
      \begin{array}{rclcrcl}
        \cL_m\ket{h,\ell,t;\theta}&=&0\,,\quad m\geq1\,,\quad&
        \cQ_\lambda\ket{h,\ell,t;\theta}&=&0\,,
        &\lambda=-\theta+p\,,\quad p=1,2,\ldots\\
        \cH_m\ket{h,\ell,t;\theta}&=&0\,,\quad m\geq1\,,&
        \cG_\nu\ket{h,\ell,t;\theta}&=&0\,,&\nu=\theta+p\,,
        \quad p=0,1,2,\ldots
      \end{array}\\
      \new\begin{array}{rcl}
        (\cH_0+\frac{\ctop}{3}\theta)\,\ket{h,\ell,t;\theta}&=&
        h\,\ket{h,\ell,t;\theta}\,,\\
        (\cL_0+\theta\cH_0+\frac{\ctop}{6}(\theta^2+\theta))
        \,\ket{h,\ell,t;\theta}&=&\ell\,\ket{h,\ell,t;\theta}
      \end{array}
    \end{array}
    \label{genhw}
  \end{equation}
  is called a {\it twisted massive \hw{} state\/}.
\end{dfn}

\subsection{Branching of extremal diagrams and the topological \hw{}
  conditions}\lvm In the general position, every arrow in the extremal
diagram of a massive Verma module can be, up to a factor, inverted by
acting with the opposite mode of the other fermion:
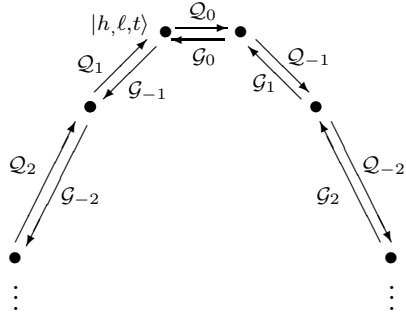
\begin{equation}
  \unitlength=1.00mm
  \begin{picture}(140,46)
    \put(50.00,10.00){
      \put(00.00,00.00){$\bullet$}
      \put(10.00,20.00){$\bullet$}
      \put(10.00,20.00){$\bullet$}
      \put(20.00,30.00){$\bullet$}
      \put(30.00,30.00){$\bullet$}
      \put(40.00,20.00){$\bullet$}
      \put(50.00,00.00){$\bullet$}
      \put(01.00,03.00){\vector(1,2){8}}
      \put(10.50,18.50){\vector(-1,-2){8}}
      \put(11.50,23.00){\vector(1,1){7}}
      \put(19.70,29.00){\vector(-1,-1){7}}
      \put(22.20,31.50){\vector(1,0){7}}
      \put(28.70,29.95){\vector(-1,0){7}}
      \put(33.00,30.00){\vector(1,-1){7}}
      \put(39.00,22.00){\vector(-1,1){7}}
      \put(43.00,19.00){\vector(1,-2){8}}
      \put(49.50,03.00){\vector(-1,2){8}}
      \put(00.00,13.00){${}_{\cQ_2}$}
      \put(07.00,07.00){${}^{\cG_{-2}}$}
      \put(09.00,26.50){${}_{\cQ_{1}}$}
      \put(16.00,21.00){${}^{\cG_{-1}}$}
      \put(23.90,33.50){${}_{\cQ_{0}}$}
      \put(24.50,25.50){${}^{\cG_{0}}$}
      \put(37.00,27.50){${}_{\cQ_{-1}}$}
      \put(32.50,22.00){${}^{\cG_{1}}$}
      \put(47.00,13.00){${}_{\cQ_{-2}}$}
      \put(41.00,07.00){${}^{\cG_{2}}$}
      \put(11.00,32.00){${}_{\ket{h_,\ell,t}}$}
      \put(00.50,-06.00){$\vdots$}
      \put(50.50,-06.00){$\vdots$}
      }
  \end{picture}
  \label{massdiagramdouble}
\end{equation}
as can be seen from the following identities:
\begin{equation}\new
  \begin{array}{rcll}
    \cQ_{-\theta}\,\ket{h,\ell,t;\theta}&=&
    2\ell\,
    \ket{h-\frac{2}{t},\ell+h-\frac{2}{t},t;\theta+1}\,,&
    \theta<0\,,\\
    \cG_{\theta-1}\,\ket{h,\ell,t;\theta}&=&
    2(\ell - h)\,
    \ket{h+\frac{2}{t},\ell-h,t;\theta-1}\,,&\theta>0\,.
  \end{array}
  \label{mapback}
\end{equation}

It follows from \req{mapback} that, as soon as one of the factors on
the right-hand sides vanishes, the respective state satisfies the
twisted {\it topological\/} \hw{} conditions~\req{twistedtophw}.  At
these `topological points', the extremal diagram branches, e.g.,
choosing~$\ell=-2(h-\frac{3}{t})$ we have
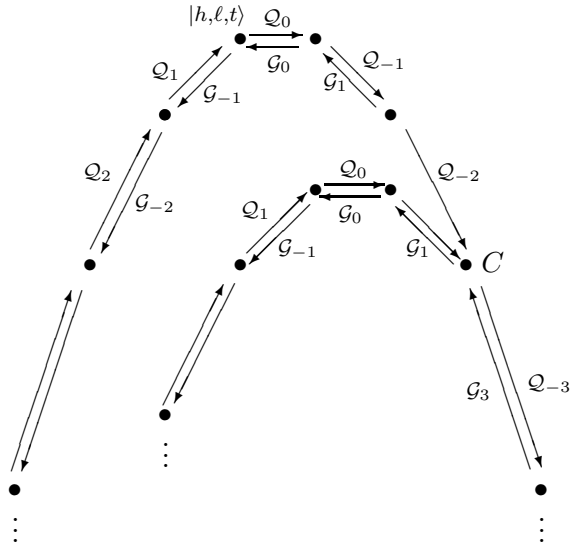
\begin{equation}
  \unitlength=1.00mm
  \begin{picture}(140,66)
    \put(30.00, 00.00){
      \put(00.00,00.00){$\bullet$}
      \put(10.00,30.00){$\bullet$}
      \put(20.00,50.00){$\bullet$}
      \put(20.00,50.00){$\bullet$}
      \put(30.00,60.00){$\bullet$}
      \put(40.00,60.00){$\bullet$}
      \put(50.00,50.00){$\bullet$}
      \put(60.00,30.00){$\bullet$}
      \put(70.00,00.00){$\bullet$}
      \put(50.00,40.00){$\bullet$}
      \put(40.00,40.00){$\bullet$}
      \put(30.00,30.00){$\bullet$}
      \put(20.00,10.00){$\bullet$}
      \put(11.00,33.00){\vector(1,2){8}}
      \put(20.50,48.50){\vector(-1,-2){8}}
      \put(00.50,03.50){\vector(1,3){8}}
      \put(09.90,27.50){\vector(-1,-3){8}}
      \put(21.50,53.00){\vector(1,1){7}}
      \put(29.70,59.00){\vector(-1,-1){7}}
      \put(32.20,61.50){\vector(1,0){7}}
      \put(38.70,59.95){\vector(-1,0){7}}
      \put(43.00,60.00){\vector(1,-1){7}}
      \put(49.00,52.00){\vector(-1,1){7}}
      \put(53.00,49.00){\vector(1,-2){8}}
      \put(63.00,28.00){\vector(1,-3){8}}
      \put(69.50,03.80){\vector(-1,3){8}}
      \put(59.30,31.00){\vector(-1,1){7.7}}
      \put(52.50,39.40){\vector(1,-1){7.7}}
      \put(42.10,41.60){\vector(1,0){8}}
      \put(49.55,39.98){\vector(-1,0){8}}
      \put(31.80,32.50){\vector(1,1){8}}
      \put(40.20,39.00){\vector(-1,-1){8}}
      \put(30.20,28.50){\vector(-1,-2){8}}
      \put(20.70,12.30){\vector(1,2){8}}
      \put(10.00,43.00){${}_{\cQ_2}$}
      \put(17.00,37.00){${}^{\cG_{-2}}$}
      \put(19.00,56.50){${}_{\cQ_{1}}$}
      \put(26.00,51.50){${}^{\cG_{-1}}$}
      \put(33.90,63.50){${}_{\cQ_{0}}$}
      \put(34.50,55.50){${}^{\cG_{0}}$}
      \put(47.00,58.00){${}_{\cQ_{-1}}$}
      \put(42.00,53.00){${}^{\cG_{1}}$}
      \put(57.00,43.00){${}_{\cQ_{-2}}$}
      \put(69.00,15.00){${}_{\cQ_{-3}}$}
      \put(61.00,12.00){${}^{\cG_{3}}$}
      \put(53.00,31.00){${}^{\cG_{1}}$}
      \put(44.00,35.50){${}^{\cG_{0}}$}
      \put(44.00,43.50){${}_{\cQ_{0}}$}
      \put(31.00,38.00){${}_{\cQ_{1}}$}
      \put(36.00,31.00){${}^{\cG_{-1}}$}
      \put(24.00,62.00){${}^{\ket{h,\ell,t}}$}
      \put(63.00,30.00){$C$}
      \put(00.50,-06.00){$\vdots$}
      \put(70.50,-06.00){$\vdots$}
      \put(20.50,04.00){$\vdots$}
      }
  \end{picture}
  \label{branchdiag}
\end{equation}

\bigskip

A crucial fact is that, once we are on the inner parabola, we can
never leave it: none of the operators from the $\N2$ algebra maps onto
the remaining part of the big parabola from the small one.  In other
words, the inner diagram in \req{branchdiag} corresponds to an $\N2$
{\it subrepresentation\/}.

These simple observations are summarized as a theorem on a class of
singular vectors in the massive Verma modules. By iteratively applying
relations \req{mapback}, we arrive at the `if' statement of the
following Theorem:
\begin{thm}\label{charged:thm}
  A massive Verma module $\mU_{h,\ell,t}$ contains a twisted
  topological Verma submodule if and only if $\ell=\theel_{\rm
    ch}(r,h,t)$, where
  \begin{equation}
    \theel_{\rm ch}(r,h,t)=r(h+\frac{r-1}{t})\,,\quad r\in\oZ;
    \label{3rdconditions}
  \end{equation}
  The corresponding singular vector reads
  \begin{equation}
    \ket{E(r,h,t)}_{\rm ch}=\left\{\kern-4pt\new\begin{array}{ll}
        \cQ_{r}\,\ldots\,\cQ_0\,\ket{h,\theel_{\rm ch}(r,h,t),t}&r\leq0\,,\\
        \cG_{-r}\,\ldots\,\cG_{-1}\,\ket{h,\theel_{\rm ch}(r,h,t),t}\,,&r\geq1
      \end{array}\right.
    \label{thirdE}
  \end{equation}
  It satisfies the twisted topological \hw{} conditions
  \req{twistedtophw} with $\theta=-r$.
\end{thm}
These singular vectors are from the `charged' series of~\cite{[BFK]}.
To be more precise, the charged singular vectors of Ref.~\cite{[BFK]}
are the top-level representatives of the extremal diagram of the
submodule built on the charged singular vector. This representative
satisfies the highest-weight conditions~\req{tophw} and is constructed
as a descendant of~\req{thirdE} as follows:
\begin{equation}
  \ket{s(r,h,t)}_{\rm ch}=\left\{
    \new\begin{array}{ll}
      \cG_0\,\ldots\,\cG_{-r-1}\,\cQ_{r}\,\ldots\,\cQ_0\,
      \ket{h,\theel_{\rm ch}(r,h,t),t}\,,
      &r\leq-1\,,\\
      \cQ_1\,\ldots\,\cQ_{r-1}\,\cG_{-r}\,\ldots\,\cG_{-1}\,
      \ket{h,\theel_{\rm ch}(r,h,t),t}\,,
      &r\geq1\,.
    \end{array}\right.
  \label{thirdplus}\label{thirdminus}
\end{equation}
In the generic situation, the same submodule is generated by either of
the representatives -- the one that satisfies topological \hw{}
conditions and the top-level one. In some degenerate cases, however,
the top-level representative may not generate the entire submodule, in
which case subsingular vectors would have to be considered,
see~\cite{[ST4]} for the details. On the other hand, the topological
representative (which we will simply refer to as the charged singular
vector) generates the maximal submodule.

\medskip

An important remark is in order regarding the structure of submodules
in $\N2$ Verma modules. As can be seen from~\req{branchdiag}, the
submodule generated from $C$ is a twisted topological Verma module,
its characteristic property being that the state $C$ satisfies
stronger annihilation conditions than the other states in the extremal
diagram; in the language of extremal diagrams, this state is
represented by a `cusp'.  A similar `cusp' necessarily exists in the
extremal diagram of any submodule of a topological Verma module.
Consider, again, the diagram~\req{topdiag} and a submodule generated
from a twisted topological singular vector (the `cusp' in the extremal
diagram of the submodule):
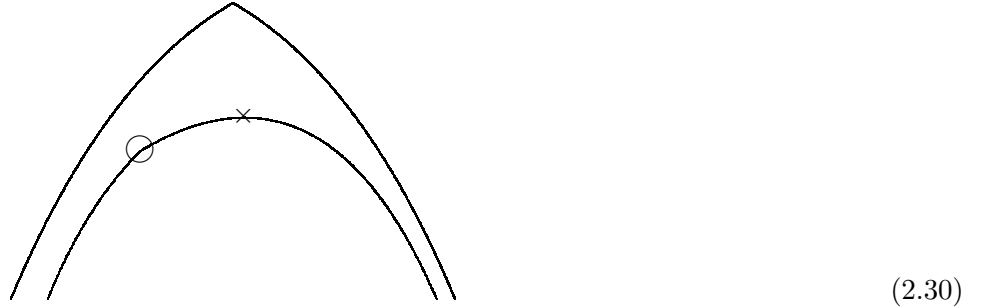
\begin{equation}
  \unitlength=.7pt
  \begin{picture}(400,160)
    \put(40,0){
      \bezier{800}(0,0)(50,120)(120,160)
      \bezier{800}(120,160)(190,120)(240,0)
      \bezier{200}(20,0)(40,50)(70,80)
      \bezier{600}(70,80)(170,140)(230,0)
      \put(120,95){$\times$}
      \put(70,81){\circle{14}}
      }
  \end{picture}
  \label{topsubmodules}
\end{equation}
When such a submodule is viewed as being generated from the {\it
  top-level\/} vector (marked with a cross in the diagram), it might
be taken for a massive Verma module. This is the reason behind the
confusion existing in the literature as regards the embedding diagrams
and other properties of submodules (and, in fact, singular vectors) in
$\N2$ Verma modules.

A similar situation exists with submodules generated from the charged
singular vectors in massive Verma modules: any such submodule is a
twisted topological Verma module, which is concealed by the fact that
the top-level representative of the extremal diagram of the submodule
satisfies the massive, not topological, \hw{} conditions; it is only
some distance along the extremal diagram that one encounters a state
satisfying twisted {\it topological\/} \hw{} conditions
\req{twistedtophw} with some $\theta$. However, the presence of such a
state (a `cusp' in the diagram, shown with a circle) drastically
changes the nature of the submodule (e.g., the topological and the
massive Verma modules have different characters, etc.).

\subsection{`Massive' singular vectors}\lvm
As we have seen, the ``charged'' series of singular vectors in massive
Verma modules are constructed very easily, in fact they immediately
follow from the analysis of extremal diagrams. This does not exhaust
all singular vectors in massive Verma modules however.

Our strategy in finding explicit expressions for all singular vectors
in $\N2$ Verma modules will be to develop the observation that a
singular vector appears as soon as there is a topological point in the
diagram~\req{massdiagramdouble}. We extend the notion of extremal
vectors to arbitrary complex $\theta$, then any diagram
\req{massdiagramdouble} can be considered as having a branching point,
albeit a `non-integral' one. This essentially reduces the problem of
constructing the general massive singular vectors to constructing
topological singular vectors.

However, we should first of all define precisely which singular
vectors are massive (i.e., generate massive Verma modules) and which
are topological (i.e., generate twisted topological Verma modules).
In the definition of massive singular vectors, we have to ensure the
following three points: that a given state, indeed, generate a {\it
  submodule}, and that this not be a topological submodule.  On the
other hand, a given submodule can be generated from different vectors,
all of which are then referred to as {\it representatives\/} of the
singular vector (in fact, of the extremal diagram of the submodule).
\begin{dfn}
  Let $\ket{Y}$ be a state in an $\N2$ Verma module that satisfies
  twisted massive \hw{} conditions with some $\theta\in\oZ$. Then
  $\ket{X}$ is said to be a dense $\cG/\cQ$-descend\-ant of~$\ket{Y}$
  if either
  $$
  \ket{X}=\alpha\,
  \cG_{\theta-N}\,\ldots\,\cG_{\theta-1}\,\ket{Y}\,,\quad N\in\oN\,,
  $$
  or
  $$
  \ket{X}=\alpha\,
  \cQ_{-\theta-M}\,\ldots\,\cQ_{-\theta}\,\ket{Y}\,,\quad M\in\oN_0\,.
  $$
  for some $\alpha\in\oC\setminus\{0\}$.
\end{dfn}
\begin{dfn}\label{defsingmass}
  A representative of a massive singular vector in the massive Verma
  module $\mU_{h,\ell,t}$ is any element of $\mU_{h,\ell,t}$ such that
  \begin{itemize}
    \addtolength{\parskip}{-6pt}
    
  \item[i)] it satisfies twisted massive \hw{} conditions, i.e., it is
    annihilated by the operators $\cL_m$, $\cH_m$, $m\in\oN$, \ 
    $\cQ_\lambda$, $\lambda\in-\theta+\oN$, and $\cG_\nu$,
    $\nu=\theta+\oN_0$ with some $\theta\in\oZ$,
    
  \item[ii)] none of its dense $\cG/\cQ$-descend\-ants vanishes,
    
  \item[iii)] the \hw{} state $\ket{h,\ell,t}$ is {\it not\/} one of
    its descendants.

  \end{itemize}
\end{dfn}

In order to choose a representative of the extremal subdiagram, we,
again, have to fix its relative charge. It will be useful to choose
the representatives $\ket{S(r,s,h,t)}^-$ and $\ket{S(r,s,h,t)}^+$ such
that $({\rm charge},{\rm level})=(-rs,\half(rs+1)(rs+2)-1)$ and
$(rs,\half rs(rs+1))$, respectively.  In what follows, we construct
these singular vectors explicitly and discuss their properties.
Another choice could be the {\it top-level representative\/} that
satisfies the $\theta=0$ case of \hw{} conditions \req{integertheta},
namely (with $\approx$ standing for the equalities that hold on the
\hw{} state)
\begin{equation}
  \label{massivehw}
  \cQ_{\geq1}\approx\cG_{\geq0}\approx
  \cL_{\geq1}\approx\cH_{\geq1}\approx0\,,
\end{equation}
with $({\rm charge},{\rm level})=(0,rs)$. In the general position, all
the three vectors $\ket{S(r,s,h,t)}^+$, $\ket{S(r,s,h,t)}^-$, and the
top-level representative generate the same submodule, and it is
therefore irrelevant which one to select. In many degenerate cases,
however, the top-level representative generates only a {\it
  sub\/}module of the module generated from $\ket{S(r,s,h,t)}^+$ or
$\ket{S(r,s,h,t)}^-$ (see~\cite{[ST4]} for the classification of
degenerations of $\N2$ Verma modules).

Applying to the above definitions the spectral flow transform would
produce the necessary modifications for the twisted modules
$\smU_{h,\ell,t;\theta}$. We will thus explicitly construct singular
vectors only in untwisted massive Verma modules.
\begin{lemma}[\cite{[BFK]}]
  A massive Verma submodule exists in $\mU_{h,\ell,t}$ if and only if
  $\ell=\theel(r,s,h,t)$, where
  \begin{eqnarray}
    \theel(r,s,h,t)&=&
    -\frac{t}{4}(h-\hminus(r,s,t))(h-\hplus(r,s+1,t))\,,
    \label{theell}\\
    (r,s,h,t)&\in&(\oN\times\oN\times\oC\times\oC)\nonumber
  \end{eqnarray}
\end{lemma}

\section{Continued extremal states, and continued products of
  fermionic generators\label{subsec:AlgRules}}\lvm In this section, we
extend the definition of twisted \hw{} vectors to the case of
non-integral (in fact, complex) twist parameter $\theta$. \ Thus the
{\it generalized topological \hw{} vectors\/} $\kettop{h,t;\theta}$
and {\it generalized massive \hw{} vectors} $\ket{h,\ell,t;\theta}$
are defined by the same annihilation conditions as the twisted \hw{}
vectors, Eqs.~\req{twistedtophw}--\req{Cartantheta2} and
\req{integertheta}--\req{genhw} respectively, {\it with arbitrary\/}
$\theta\in\oC$.  Accordingly, {\it generalized topological Verma
  modules\/} $\smV_{h,t;\theta}$ and {\it generalized massive Verma
  modules\/} $\smU_{h,\ell,t;\theta}$ are freely generated from the
generalized \hw{} states by the respective creation operators, where
it is understood that any `$\theta$'-module is considered over the
algebra $\cA_\theta$, which is the image of the algebra
\req{topalgebra} under the spectral flow transform $\cU_\theta$; thus
the modes of $\cG$ are running over $\theta+n$, $n\in\oZ$, and those
of $\cQ$, over $-\theta+n$, $n\in\oZ$ (of which $\cG_{\mu}$ with
$\mu\in\theta-1-\oN_0$ and $\cQ_{\nu}$ with $\nu\in-\theta-\oN_0$ are
the creation operators in the generalized massive Verma modules).
Obviously, the generalized Verma modules become the twisted Verma
modules whenever $\theta\in\oZ$.

In the thus defined generalized modules, one considers the extremal
states, e.g.
$\cG_{\theta-N}\,\ldots\,\cG_{\theta-1}\cdot\,\ket{h,\ell,t;\theta}$,
$\theta\in\oC$. These are further continued in
$\theta-N\leadsto\theta'$, with $\theta'\in\oC$ not necessarily
differing from $\theta$ by an integer. Such generalized extremal
states are in fact elements of another generalized Verma module,
namely the one characterized by the parameter $\theta'$ (therefore, in
accordance with the above, the states are acted upon with $\cG_\mu$
with $\mu\in\theta'+\oZ$, and $\cQ_\nu$, $\nu\in-\theta'+\oZ$).

Rather than working with states, we introduce the {\it operators\/}
$q(-\theta-n,-\theta)=
\cQ_{-\theta-N}\,\cQ_{-\theta-N+1}\,\ldots\,\cQ_{-\theta}$ \ and \ 
$g(\theta-N,\theta)=
\cG_{\theta-N}\,\cG_{\theta-N+1}\,\ldots\,\cG_{\theta}$ and then
continue them to arbitrary complex arguments,
\begin{equation}
  g(\theta',\theta)\quad {\rm and} \quad q(\theta',\theta)\,,
  \quad \theta',\theta\in\oC\,,
  \label{gandq}
\end{equation}
by postulating their algebraic properties in such a way that, whenever
$\theta-\theta'\in\oN$, the properties become those of the above
products of modes.  This approach of dealing with relative objects
(operators) rather than with the vectors on which they act, proves to
be very fruitful because the new operators possess rich algebraic
properties.

\bigskip

The fact that the new operators $g(a,b)$ and $q(a,b)$ can be thought
of as an extension of the products of modes to a non-integral (in
fact, complex) number of factors is formalized as follows:

Define the {\it length\/} of \ $g(a,b)$ or $q(a,b)$ \ as \ $b-a+1$.
Then, the `tautological' property of the new operators is

\paragraph{Positive integral length reduction.} Whenever the length is
a non-negative integer, the operator $g(a,b)$ or $q(a,b)$ becomes the
product of the corresponding modes:
\begin{equation}
  g(a,b)=\prod_{i=0}^{L-1}\cG_{a+i}\,,\quad q(a,b)=\prod_{i=0}^{L-1}\cQ_{a+i}\,,
  \quad{\rm iff}\quad L\equiv b-a+1=0,1,2,\ldots
  \label{integrallength}
\end{equation}
(in the case $L=0$ the product evaluates as 1). By definition, these
products are ordered as $\ldots\cG_{a}\cG_{a+1}\ldots$.

\paragraph{Gluing rules.\label{GlueII}}
\begin{equation}\new
  \begin{array}{rcl}
    g(a,b-1)\,g(b,\theta-1)\,\ket{\theta}_{g}&=&
    g(a,\theta-1)\,\ket{\theta}_{g}\,\\
    q(a,b-1)\,q(b,-\theta-1)\,\ket{\theta}_{q}&=&
    q(a,-\theta-1)\,\ket{\theta}_{q}\,
  \end{array}
  \quad a,b,\theta\in\oC\,.
  \label{Glue}
\end{equation}
where $\ket{\theta}_{g}$ is any state that satisfies
$\cG_{\theta+n}\ket{\theta}_{g}=0$ for $n\in\oN_0$, and
$\ket{\theta}_{q}$, similarly, satisfies
$\cQ_{-\theta+n}\ket{\theta}_{q}=0$ for $n\in\oN_0$.

\paragraph{Under the spectral flow transform} \req{U}, the
operators $g(a,b)$ and $q(a,b)$ behave in the manner inherited from
the behaviour of the products~\req{integrallength}:
\begin{equation}
  {\cal U}_\theta:\new
  \begin{array}{rcl}g(a,b)&\mapsto&g(a+\theta,b+\theta)\,,\\
    q(a,b)&\mapsto&q(a-\theta,b-\theta)\,.
  \end{array}
  \label{Ugq}
\end{equation}

\bigskip

Further properties of the new operators originate in the fact that,
the $\N2$ generators $\cQ$ and $\cG$ being fermions, they satisfy the
vanishing formulae such as, e.g.,
$$
\cG_{n}\cdot\prod_{i=a}^{a+N}\cG_{i}=0\,,\quad N\in\oN_0\,,\quad
a\leq n\leq a+N\,.
$$

\paragraph{`Pauli principle'$\!$.}
\begin{equation}
  \cG_a\,g(b,c)=0\,,\qquad
  \cQ_a\,q(b,c)=0\,,\quad a-b\in\oN_0\quad\mbox{\rm and}\quad
  (a-c\not\in\oN\quad\mbox{\rm or}\quad b-c-1\in\oN)
  \label{rightkill}
\end{equation}
Similarly, the `left-handed' annihilation properties are expressed by
the relations
\begin{equation}
  g(a,b)\,\cG_c = 0\,,\qquad q(a,b)\,\cQ_c = 0\,,
  \quad b-c\in\oN_0\quad\mbox{\rm and}\quad
  (a-c\not\in\oN\quad\mbox{\rm or}\quad a-b-1\in\oN)\,.
  \label{leftkill}
\end{equation}

\medskip

We will also need some vanishing conditions with respect to the
bosonic operators $\cL_{\geq1}$ and $\cH_{\geq1}$. These can be
derived from the following basic commutation relations for the
continued operators:

\paragraph{Positive-moded bosons.} The bosonic generators $\cL_p$
and $\cH_p$ with $p=1,2,\ldots$, commute with operators \req{gandq} as
\begin{equation}\new
  \begin{array}{rcl}
    \left[\cK_n,\, g(a,b)\right]\kern-4pt&=&\kern-4pt
    \sum_{p=0}^{d(n,a,b)}g(a,b-p-1)
    \left[\cK_n\,,\,\cG_{b-p}\right]\,\cG_{b-p+1}\ldots \cG_b\,,
    \quad n\in\oN\,,\\
    \left[\cK_n,\, q(a,b)\right]\kern-4pt&=&\kern-4pt
    \sum_{p=0}^{d(n,a,b)}q(a,b-p-1)
    \left[\cK_n\,,\,\cQ_{b-p}\right]\,\cQ_{b-p+1}\ldots \cQ_b\,,
  \end{array}
  \label{LRight}
\end{equation}
where $\cK=\cL$ or $\cH$ and
\begin{equation}
  d(n,a,b)=\left\{\new\begin{array}{ll}
      b-a\,,&n-b+a\in\oN_0\quad\mbox{\rm and}\quad b-a+1\in\oN_0\,,\\
      n-1\,,  &{\rm otherwise}
    \end{array}\right.
  \label{ddd}
\end{equation}
All the commutators $[\cK_n\,,\,\cG_{b-l}]$ and
$[\cK_n\,,\,\cQ_{b-l}]$ are to be taken from \req{topalgebra1}.  In
the case of a positive integral length $b-a+1\in\oN$, the relations
\req{LRight} turn into identities in the universal enveloping of the
$\N2$ algebra. In fact, Eqs.~\req{LRight} are the algebraic
continuation of those identities to the case of complex $a$ and $b$.
The main point is that, even though the length $b-a+1$ may not be an
integer, there is always an integral number of terms on the RHS of
\req{LRight} (this {\it is\/} reconciled with a naive application of
the Leibnitz rule, according to which $\cK_n$ should be commuted with
each of the modes $\cG_\mu$ `virtually contained' in $g(a,b)$;
according to the Pauli principle, the application of the Leibnitz rule
is restricted to the last $n$ modes of the $\cG_\mu$).

\paragraph{Eigenvalues.} Application of the operators $g$ and $q$
changes the eigenvalues of $\cL_0$ and $\cH_0$.  The effect can be
expressed by the commutation relations
\begin{equation}\new
  \begin{array}{rclcrcl}
    {[}\cL_0,\,g(a,b)]&=&-\half(a+b)(b-a+1)\,g(a,b)\,,&{}&
    [\cH_0,\,g(a,b)]&=&(b-a+1)\,g(a,b)\,,\\
    {[}\cL_0,\,q(a,b)]&=&-\half(a+b)(b-a+1)\,q(a,b)\,,&{}&
    [\cH_0,\,q(a,b)]&=&(-b+a-1)\,q(a,b)\,.
  \end{array}
  \label{L0H0}
\end{equation}
Underlying the commutation properties~\req{L0H0} is the same intuitive
idea as above, that $g(a,b)$ represents the states from $a$ to $b$
filled with fermions $\cG_\mu$. Namely, the eigenvalues on the RHSs of
\req{L0H0} are simply the number of such fermionic factors in the case
of $\cH_0$ and (minus) the sum of their modes in the case of $\cL_0$.

\paragraph{Positive-moded fermions.} Annihilation properties with
respect to the fermionic generators can be arrived at as follows.  For
a negative integral $\theta$ we have a relation in the Verma module
$\mU_{h,\ell,t}$
\begin{equation}\new
  \begin{array}{rcl}
    \cQ_{-\theta+n}\,g(\theta,-1)\,
    \ket{h,\ell,t}&=&
    \sum_{i=0}^{-\theta-1}(-1)^ig(\theta,\theta+i-1)\,
    \{\cQ_{-\theta+n},\,\cG_{\theta+i}\}\,
    g(\theta+i+1,-1)\,
    \ket{h,\ell,t}\,,\\
    {}&{}&{}n\in\oN_0\,,\quad\theta\in-\oN\,.
  \end{array}
  \label{cross}
\end{equation}
In this form, the formula does not continue to $\theta\in\oC$.
However, when we insert the commutators
$\{\cQ_{-\theta+n},\,\cG_{\theta+i}\}$ from \req{topalgebra1} and
evaluate the resulting modes of $\cL$ and $\cH$ according to
\req{LRight}, we see that \req{cross} vanishes for $n\geq1$. We
continue this to
\begin{equation}
  \cQ_{-\theta+n}\,g(\theta,-1)\,
  \ket{h,\ell,t}=0\,,\quad\theta\in\oC\,,\quad n\in\oN\,,
  \label{(3.9)}
\end{equation}
while in the case of $n=0$ we are left with
\begin{equation}\new\begin{array}{rclr}
    \cQ_{-\theta}\,g(\theta,-1)\,\ket{h,\ell,t}&=&
    \cQ_{-\theta}\,\cG_{\theta}\,
    g(\theta+1,-1)\,\ket{h,\ell,t}&{\rm by\ \req{Glue}}\\
    {}&=&\{\cQ_{-\theta},\,\cG_{\theta}\}\,g(\theta+1,-1)
    \,\ket{h,\ell,t}&{\rm by}\ \req{(3.9)}\\
    {}&=&{}\bigl(2\cL_0+2\theta\cH_0+\frac{\ctop}{3}(\theta^2+\theta)\bigr)
    \,g(\theta+1,-1)\,\ket{h,\ell,t}&{\rm by}\
    \req{topalgebra1}
  \end{array}\end{equation}
Recalling now Eq.~\req{L0H0}, we arrive at
\begin{equation}
  \cQ_{-\theta}\,g(\theta,-1)\,\ket{h,\ell,t}=
  2(\ell+\theta h-\frac{1}{t}({\theta}^2+\theta))
  \,g(\theta+1,-1)\,\ket{h,\ell,t}
  \label{cross2new}
\end{equation}

Similarly, for the $q$-operators, we have the following properties:
\begin{equation}\new
  \begin{array}{rcl}
    \cG_{\theta+n}\,q(-\theta,0)\,\ket{h,\ell,t}&=&0\,,\quad n\in\oN\,,\\
    \cG_{\theta}\,q(-\theta,0)\,\ket{h,\ell,t}&=&
    2(\ell+\theta h-\frac{1}{t}(\theta^2+\theta))q(-\theta+1,0)\,\ket{h,\ell,t}\,.
  \end{array}
  \label{massiveanih}
\end{equation}
The formulae \req{(3.9)}, \req{cross2new}, \req{massiveanih} are now
viewed as continued to~$\theta\in\oC$.

Similarly to \req{cross2new} and \req{massiveanih}, we derive the
following relations for integral parameters and then postulate for all
$\theta,\theta'\in\oC$:
\begin{equation}
  \cQ_{-\theta'}\,g(\theta',\theta-1)\,\ket{h,t;\theta}_{\rm top}=
  2(\theta'-\theta)(h+\frac{1}{t}(\theta-\theta'-1))
  \,g(\theta'+1,\theta-1)\,\ket{h,t;\theta}_{\rm top}
  \label{cross2}
\end{equation}
and
\begin{equation}
  \cG_{\theta'}\,q(-\theta',-\theta-1)\,\ket{h,t;\theta}_{\rm top}=
  2(\theta'-\theta)(h+1+\frac{1}{t}(\theta-\theta'-1))
  \,q(\theta'+1,\theta-1)\,\ket{h,t;\theta}_{\rm top}
  \label{cross1}
\end{equation}

\bigskip

We have thus given the list of those algebraic properties of the $q$
and $g$ operators that are related to the highest-weight/annihilation
conditions.  The following property is of a somewhat different spirit,
but it can be derived in a direct analogy with \req{LRight}:

\paragraph{Negative-moded bosons.} In addition to the various types
of the highest-weight conditions, which apply to essentially the
`positive'-moded generators, we will also need to commute the
negative-moded $\cH$ and $\cL$ operators through $q(a,b)$ and
$g(a,b)$. The corresponding formulae read
\begin{equation}\new
  \begin{array}{rcl}
    \left[g(a,b),\,\cK_p\right]\kern-4pt&=&\kern-6pt
    \sum_{l=0}^{d(-p,a,b)}\cG_a\ldots \cG_{a+l-1}
    \left[\cG_{a+l}\,,\,\cK_p\right]\,g(a+l+1,b)\,,\\
    \left[q(a,b),\,\cK_p\right]\kern-4pt&=&\kern-6pt
    \sum_{l=0}^{d(-p,a,b)}\cQ_a\ldots \cQ_{a+l-1}
    \left[\cQ_{a+l}\,,\,\cK_p\right]\,q(a+l+1,b)\,,
  \end{array}
  \label{LLeft}
\end{equation}
where $p\in-\oN$ and $d(p,a,b)$ is given by formula~\req{ddd}.  As
before, $\cK=\cH$ or $\cL$.

\paragraph{Highest-weight conditions.} The above annihilation
properties~\req{LRight}, \req{(3.9)}, and \req{massiveanih} allow us
to formulate the following assertion regarding the \hw{} properties of
continued states $g(\theta,-1)\,\ket{h,\ell,t}$ and
$q(-\theta,0)\,\ket{h,\ell,t}$, which turn out to be the generalized
\hw{} states:\pagebreak[3]

\begin{lemma}\label{lemma:HW}\mbox{}\nopagebreak
  \begin{enumerate}
    \addtolength{\parskip}{-6pt}
    
  \item The objects $g(\theta,-1)\,\ket{h,\ell,t}$ and
    $q(-\theta,0)\,\ket{h,\ell,t}$ satisfy the following annihilation
    conditions:
    \begin{equation}\new
      \begin{array}{rcl}
        \cL_m\,g(\theta,-1)\,\ket{h,\ell,t}&=&0\,,\quad
        \cH_m\,g(\theta,-1)\,\ket{h,\ell,t}=0\,,\quad m\in\oN\,,\\
        \cG_a\,g(\theta,-1)\,\ket{h,\ell,t}&=&0\,,\quad a\in\theta+\oN_0\,,\\
        \cQ_a\,g(\theta,-1)\,\ket{h,\ell,t}&=&0\,,\quad a\in-\theta+\oN\,.
      \end{array}
      \label{rightkillg}
    \end{equation}
    and
    \begin{equation}\new
      \begin{array}{rcll}
        \cL_m\,q(-\theta,0)\,\ket{h,\ell,t}&=&0\,,&
        \cH_m\,q(-\theta,0)\,\ket{h,\ell,t}=0\,,\quad m\in\oN\,,\\
        \cG_a\,q(-\theta,0)\,\ket{h,\ell,t}&=&0\,,&a\in\theta+\oN\,,\\
        \cQ_a\,q(-\theta,0)\,\ket{h,\ell,t}&=&0\,,&a\in-\theta+\oN_0\,.
      \end{array}
      \label{rightkillq}
    \end{equation}
    
  \item In the topological case, we have
    \begin{equation}\new
      \begin{array}{rclrcll}
        \cG_a\,g(\theta',\theta-1)\,\ket{h,t;\theta}_{\rm top}&=&0\,,&
        \cQ_a\,q(\theta',-\theta-1)\,\ket{h,t;\theta}_{\rm
          top}&=&0\,,& a-\theta'\in\oN_0\,,\\
        \cQ_a\,g(\theta',\theta-1)\,\ket{h,t;\theta}_{\rm top}&=&0\,,&
        \cG_a\,q(\theta',-\theta-1)\,\ket{h,t;\theta}_{\rm
          top}&=&0\,,& a+\theta'\in\oN\,\\
        \cL_m\,g(\theta',\theta-1)\,\ket{h,t;\theta}_{\rm top}&=&0\,,&
        \cH_m\,q(\theta',-\theta-1)\,\ket{h,t;\theta}_{\rm
          top}&=&0\,,& m\in\oN\,,\\
        \cH_m\,g(\theta',\theta-1)\,\ket{h,t;\theta}_{\rm top}&=&0\,,&
        \cL_m\,q(\theta',-\theta-1)\,\ket{h,t;\theta}_{\rm
          top}&=&0\,,& m\in\oN\,.
      \end{array}
      \label{LHkill}\label{rightkill2}
    \end{equation}
  \end{enumerate}
\end{lemma}

Now, we can prove the following
\begin{thm}
  Up to a normalization factor, the objects
  $g(\theta,-1)\,\ket{h,\ell,t}$ and $q(-\theta,0)\,\ket{h,\ell,t}$
  represent generalized massive highest-weight vectors with the
  following parameters:
  \begin{equation}\new
    \begin{array}{rcl}
      \ket{h',\ell',t;\theta'}\kern-4pt&\sim&\kern-4pt
      g(\theta',\theta-1)\,\ket{h,\ell,t;\theta}\,,\\
      h'\kern-4pt&=&\kern-4pt h+\frac{2}{t}(\theta-\theta')\,,\\
      \ell'\kern-4pt &=&\kern-4pt
      \ell+(\theta'-\theta)(h-\frac{1}{t}(\theta'-\theta+1))
    \end{array}
    \label{gmapextremal}\label{hnew}
  \end{equation}
  and
  \begin{equation}\new
    \begin{array}{rcl}
      \ket{h'',\ell'',t;\theta'+1}\kern-4pt &\sim&\kern-4pt
      q(-\theta',-\theta)\ket{h,\ell,t;\theta}\,,\\
      h''\kern-4pt &=&\kern-4pt h +\frac{2}{t}(\theta-\theta'-1)\,,\\
      \ell''\kern-4pt &=&\kern-4pt
      \ell+(\theta'-\theta + 1)(h-\frac{1}{t}(\theta'-\theta+2))
    \end{array}
    \label{lnewq}\label{qmapextremal}
  \end{equation}
\end{thm}
Indeed, the annihilation properties follow by applying \req{Ugq} to
the relations of the previous Lemma. The actual parameters of the
states follows by a direct calculation, which we demonstrate for~$h'$.
{}From \req{genhw}, we have
$$
\cH_0\ket{h',\ell',t;\theta'}=
(h'-\frac{\ctop}{3}\theta')\ket{h',\ell',t;\theta'}\,.
$$
On the LHS of~\req{gmapextremal}, we evaluate $\cH_0$ using
\req{L0H0}, whence~\req{hnew} follows.  Note also that whenever
$\ell+(\theta'-\theta) h - \frac{1}{t}((\theta'-\theta)^2 + \theta' -
\theta) = 0$, Eqs.~\req{massiveanih} also allow us to show that, in
addition to~\req{lnewq},
\begin{equation}
  q(-\theta',-\theta)\ket{h,\ell,t;\theta}\sim
  \kettop{h + \frac{2}{t}(\theta-\theta')-1,t;\theta'}\,.
  \label{htopnew}
\end{equation}

\paragraph{Transitivity identities.}
We saw in the previous section that in the general positions, the
arrows in the extremal diagrams can be inverted by the action of the
`opposite' fermion.  The following formulae formalize the procedure of
`canceling' several fermionic generators from the left of the $g$ and
$q$ operators
\begin{enumerate}
  \addtolength{\parskip}{-6pt}
\item for the twisted topological \hw{} vectors,
  \begin{equation}
    g(\theta'+1,\theta-1)\,\ket{h,t;\theta}_{\rm top}=
    {1\over\Lambda_g(\theta,\theta',h,t)}\,
    \cQ_{-\theta'}\,g(\theta', \theta-1)\,\ket{h,t;\theta}_{\rm top}
    \label{gtrans}
  \end{equation}
  where
  \begin{equation}
    \Lambda_g(\theta,\theta',h,t)=
    2(\theta'-\theta)(h+\frac{1}{t}(\theta-\theta'-1))\,,
    \label{Lambdagtrans}
  \end{equation}
  and
  \begin{equation}
    q(-\theta'+1,-\theta-1)\,\ket{h,t;\theta}_{\rm top}=
    {1\over\Lambda_q(\theta,\theta',h,t)}\,
    \cG_{\theta'}\,q(-\theta',-\theta-1)\,\ket{h,t;\theta}_{\rm
      top}\label{qtrans}
  \end{equation}
  where
  \begin{equation}
    \Lambda_q(\theta,\theta',h,t)=
    2(\theta'-\theta)(h+1+\frac{1}{t}(\theta-\theta'-1))\,.
    \label{Lambdaqtrans}
  \end{equation}
  
\item for the massive \hw{} vectors,
  \begin{equation}
    g(\theta+1,-1)\,\ket{h,\ell,t}=
    {1\over\Lambda(\theta,h,\ell,t)}\,
    \cQ_{- \theta}\,g(\theta, -1)\,\ket{h,\ell,t}\,,
    \label{mgtrans}
  \end{equation}
  and
  \begin{equation}
    q(-\theta+1,0)\,\ket{h,\ell,t}=
    {1\over\Lambda(\theta,h,\ell,t)}\,
    \cG_{\theta}\,q(-\theta, 0)\,\ket{h,\ell,t}\,,
    \label{mqtrans}
  \end{equation}
  with
  \begin{equation}
    \Lambda(\theta,h,\ell,t)=2(\ell+\theta h-\frac{1}{t}(\theta^2+\theta))
    \label{Lambdaboth}
  \end{equation}
  in both cases.
\end{enumerate}
It follows that `violations of transitivity' occur whenever the
corresponding $\Lambda$-factor vanishes. Then the state on the RHS of
the respective formula satisfies stronger annihilation conditions,
namely it becomes a (generalized) topological \hw{} state (while in
the general position it is just a (generalized) massive \hw{} state).
These `enhanced' \hw{} conditions occur in the following cases:
$$\new
\begin{array}{lclcl}
  \theta'=ht+\theta-1&\Leftrightarrow&
  g(\theta',\theta-1)\ket{h,t;\theta}_{\rm top}=\ket{h',t;\theta'}_{\rm top}
  &\Leftrightarrow&
  \Lambda_g(\theta,\theta',h,t)=0
  \quad\hbox{\rm in \req{gtrans}}\\
  \theta'=(h+1)t+\theta-1&\Leftrightarrow&
  q(-\theta',-\theta-1)\ket{h,t;\theta}_{\rm top}=\ket{h',t;\theta'}_{\rm top}
  &\Leftrightarrow&
  \Lambda_q(\theta,\theta',h,t)=0
  \quad\hbox{\rm in \req{qtrans}}\\
  \ell+h\theta-\frac{1}{t}(\theta^2+\theta)=0&\Leftrightarrow&
  \!\!\!\begin{array}{rcl}
    g(\theta',-1)\ket{h,t;\ell}&=&\ket{h',t;\theta'}_{\rm top}\\
    q(-\theta',0)\ket{h,t;\ell}&=&\ket{h',t;\theta'}_{\rm top}\end{array}
  &\Leftrightarrow&\Lambda(\theta,h,\ell,t)=0\,{}\!~
  \begin{array}{l}{\rm in}\ \req{mgtrans}\\
    {\rm in}\ \req{mqtrans}
  \end{array}
\end{array}
$$

\section{Algebraic constructions of $\N2$ singular
  vectors\label{sec:Constructing}}
\subsection{Constructing the topological singular
  vectors\label{topvect}}\lvm The $\N2$ singular vectors are
constructed in this section in a monomial form, as a product of the
$q$ and $g$ operators acting on the corresponding highest-weight
vector. To begin with, observe that the mappings of generalized {\it
  topological\/} \hw{} vectors by the $g$- and $q$- operators can be
combined as follows. We apply $q$,~$g$,~$q$,~\ldots, to the \hw{}
vector $\kettop{h,t}$ either as
\begin{equation}
  \smV_{h,t;\theta}\stackrel{g(\theta_1,\theta-1)}{\longrightarrow}
  \smV_{h_1,t;\theta_1}
  \stackrel{q(-\theta_2,-\theta_1-1)}{\longrightarrow}\ldots
  \label{start-g}
\end{equation}
or as 
\begin{equation}
  \smV_{h,t;\theta}\stackrel{q(-\theta_{-1},-\theta-1)}{\longrightarrow}
  \smV_{h_{-1},t;\theta_{-1}}
  \stackrel{g(\theta_{-2},\theta_{-1}-1)}{\longrightarrow}\ldots
  \label{start-q}
\end{equation}
and requiring that the generalized topological \hw{} conditions be
preserved at each step, we see that the $h$ and $\theta$ parameters
should take on the values
\begin{equation}
  \new
  \begin{array}{rcl}
    \theta_i=&\left\{
      \new
      \begin{array}{ll}
        (h-j)t-1+\theta\quad&i=2j+1\,,\\
        jt+\theta\quad&i=2j\,.\\
      \end{array}
    \right. &
  \end{array}
  \qquad
  \new
  \begin{array}{rcl}
    h_i=&\left\{
      \new
      \begin{array}{ll}
        \frac{2}{t}-h+2j\quad&i=2j+1\,,\\
        h-2j\quad&i=2j\\
      \end{array}
    \right. &
  \end{array}
  \label{thetashs}
\end{equation}

We thus take a chain of the $q$ and $g$ operators and act with it on
the topological highest-weight state $\ket{h,t}_{\rm top}$ of the
topological Verma module $\mV_{h,t}=\smV_{h,t;0}$; it follows that the
$\theta$ parameters in \req{start-g} are
\begin{equation}\new
  \begin{array}{rcl}
    \theta_i(h)=&\left\{
      \new
      \begin{array}{ll}
        (h-j)t-1\quad&i=2j+1\,,\\
        jt\quad&i=2j\,.\\
      \end{array}
    \right. &
  \end{array}
  \label{thethetas}
\end{equation}
This is in fact similar to the singular vector construction for affine
Lie algebras~\cite{[MFF]}, however there is no affine Weyl group in
the $\N2$ case, while in the affine case it is this group that governs
the exponents in the MFF formulae.  As in the case of affine Lie
algebras, the condition for a Verma module to contain a singular
vector is that after $2s-1$
``reflections''~\req{start-g}--\req{start-q}, we return to
$\mV_{h,t}$.  Consider for definiteness the case when we start from
the action of $g$, hence $i>0$:
\begin{equation}
  \begin{picture}(150,80)
    \put(79.0,70.0){$\bullet$}
    \put(78.0,77.0){$\mV_{h,t}$}
    \put(85.0,71.5){\vector(4,-1){30}}
    \put(95.0,62.0){$g$}
    \put(115.5,60.0){$\bullet$}
    \put(122.0,60.0){\vector(2,-1){30}}
    \put(130.5,48.0){$q$}
    \put(152.0,40.0){$\bullet$}
    \put(157.0,38.0){\vector(1,-2){12}}
    \put(168.0,0.0){$\vdots$}
    \put(63.0,62.0){$g$}
    \put(41.5,60.0){$\bullet$}
    \put(48.0,64.5){\vector(4,1){30}}
    \put(20.5,45.0){$q$}
    \put(5.0,40.0){$\bullet$}
    \put(11.0,45.0){\vector(2,1){30}}
    \put(-6.0,15.0){\vector(1,2){12}}
    \put(-7.0,0.0){$\vdots$}
  \end{picture}
  \label{ring}
\end{equation}
The condition for this to happen is that $\theta_{2s-1}(h)=(h-s+1)t-1$
be a negative integer, say~$-r$.  We thus rederive the
formulae~\req{hplushminus} expressing $\htop^+$ in terms of two
positive integers $r$ and $s$. The formula for $\htop^-$ is recovered
by considering a similar loop starting and ending with a $q$ operator.

One has the following Theorem:
\begin{thm}[\cite{[ST2]}]
  All singular vectors in the topological Verma module
  $\mV_{\htop^\pm(r,s,t),t}$ over the $\N2$ superconformal algebra are
  given by the explicit construction:
  \begin{eqnarray}
    \ket{E(r,s,t)}^+\kern-4pt&=&\kern-4pt
    g(-r,(s-1)t-1)\,q(-(s-1)t,r-1-t)\,\ldots{}
    g((s-2)t-r,t-1)\,q(-t,r-1-t(s-1))\nonumber\\
    {}&{}&\qquad\qquad
    {}\cdot g((s-1)t-r,-1)\,\kettop{\hplus(r,s,t),t}\,,
    \label{Tplus}\\
    \ket{E(r,s,t)}^-\kern-4pt&=&\kern-4pt
    q(-r, (s-1) t - 1)\,g(-(s-1)t, r - t - 1)\,\ldots
    q((s-2) t - r, t-1) \, g(-t, r - 1 - (s-1) t)\nonumber\\
    {}&{}&{}\qquad\qquad{}\cdot
    q((s-1) t - r, -1)\,\kettop{\hminus(r,s,t),t}
    \label{Tminus}\\
    &&r,s\in\oN\nonumber
  \end{eqnarray}
  where the factors in the first line of each formula are
  \begin{equation}
    g(-r - t - m t + s t, -1 + m t)\,
    q(-m t, r - 1 + m t - s t)\,,\qquad
    s-1\geq m\geq1
    \label{plusfact}
  \end{equation}
  and
  \begin{equation}
    q(-r - t - m t + s t, -1 + m t)\,
    g(-m t, r - 1 + m t - s t)\,,\qquad
    s-1\geq m\geq1
    \label{minusfact}
  \end{equation}
  respectively.  The $\ket{E(r,s,t)}^\pm$ singular vectors satisfy
  twisted topological \hw\ conditions with the `spectral' parameter
  $\theta=\mp r$ and are on level $rs+\half r(r-1)$ over the
  corresponding topological highest-weight state and have relative
  charge $\pm r$.
\end{thm}

In a direct analogy with the well-known affine Lie algebra
case~\cite{[MFF],[Mal]}, ``all singular vectors'' applies literally to
non-rational $t$, while for rational~$t$, a singular vector may be
given already by a subformula of Eqs.~\req{Tplus}, \req{Tminus} as
soon as that subformula (obtained by dropping several $g$- and
$q$-operators from the left) produces an element of the Verma module.

It will be shown in section \ref{subsec:Top} how the states
\req{Tplus}, \req{Tminus} can be rewritten as polynomials in the
standard creation operators in the topological Verma modules over the
$\N2$ algebra, $\cL_{\leq-1}$, $\cH_{\leq-1}$, $\cG_{\leq-1}$,
and~$\cQ_{\leq-1}$.

\bigskip

A useful reformulation of Eqs.~\req{Tplus}, \req{Tminus} is achieved
in the form of {\it recursion relations} that allow us to construct
the topological singular vector $\ket{E(r,s,t)}^\pm$, $s\geq2$, out of
those with lower $s$: The fact is that the rightmost continued
operator from \req{Tplus}, \req{Tminus} induces the mappings
$\mV_{\htop^\pm(r,s),t}\to \mV_{\htop^\mp(r,s-1),t}$, thereby
effectively decreasing the value of $s$:
\begin{equation}
  \unitlength=1pt
  \begin{picture}(490,70)
    \put(0,0){
      \vbox{
        $$
        \new\begin{array}{lcc}
          \quad \mV_{\hplus(r,s,t),t}\\
          \quad\biggl\downarrow\lefteqn{{}_{{}^{g(\theta_1(r,s,t),-1)}}}\\
          \smV_{\hminus(r,s-1,t),t;\theta_1(r,s,t)}&
          \longrightarrow&
          \mV_{\hminus(r,s-1,t),t}
        \end{array}
        \qquad
        \new\begin{array}{lcc}
          \quad \mV_{\hminus(r,s,t),t}\\
          \quad\biggl\downarrow\lefteqn{{}_{{}^{q(-\theta_{-1}(r,s,t),-1)}}}\\
          \smV_{\hplus(r,s-1,t),t;\theta_{-1}(r,s,t)}&
          \longrightarrow&
          \mV_{\hplus(r,s-1,t),t}
        \end{array}
        $$
        }
      }
    \put(122,58){\vector(3,-2){55}}
    \put(336,58){\vector(3,-2){55}}
  \end{picture}
  \label{squares}
\end{equation}
(the horizontal arrows being the spectral flow transform mappings),
where we have rewritten $\theta_{\pm1}$ from \req{thethetas} as
$$
\theta_{\pm1}(r,s,t)=\theta_{\pm1}(\htop^\pm(r,s,t))=\pm(t(s-1)-r)
$$

Thus, introducing the singular vector {\it operators\/} $\cE$ as
$$
\ket{E(r,s,t)}^\pm=\cE^\pm(r,s,t)\,\ket{\htop^\pm(r,s,t),t}_{\rm
  top},
$$
we have the recursion relations
\begin{equation}\new
  \begin{array}{rcll}
    \cE^+(r,s,t)&=&
    g(-r,(s-1)t-1)\,\cE^{-,(s-1)t-r}(r,s-1,t)\,g((s-1)t-r,-1)\,,\\
    \cE^{-}(r,s,t)&=&
    q(-r,(s-1)t-1)\,\cE^{+,-(s-1)t+r}(r,s-1,t)\,q((s-1)t-r,-1)\,.
  \end{array}
  \label{recursion}
\end{equation}
where $\cE^{\pm,\theta}$ is the spectral flow transform of $\cE^\pm$.
These relations involve the continued operators, however the
combination in which these operators appear allows us to evaluate
Eqs.~\req{recursion} in terms of the usual $\N2$ generators.  This
will be shown in Sec.~\ref{sec:Algebra}.

\medskip

A remark is in order regarding the top-level
representatives~\cite{[BFK],[Doerr2]} of topological singular vectors,
discussed in~\req{topsubmodules}. The top-level representatives
satisfy the `untwisted' annihilation conditions~\req{massivehw}.  Such
vectors are easily constructed from the above $\ket{E(r,s,t)}^\pm$ by
traveling along the extremal diagram as follows:
\begin{equation}\new
  \begin{array}{rcl}
    \ket{s(r,s,t)}^+&=&\cQ_{0}\,\ldots\,\cQ_{r-1}\,\ket{E(r,s,t)}^+\,,\\
    \ket{s(r,s,t)}^-&=&\cG_{0}\,\ldots\,\cG_{r-1}\,\ket{E(r,s,t)}^-\,.
  \end{array}
  \label{ST}
\end{equation}
They have relative charge zero and level $rs$ with respect to the
corresponding topological highest-weight state
$\kettop{\hplusminus(r,s,t),t}$. However, there are two factors that
make working with $\ket{s(r,s,t)}^\pm$ quite complicated. Firstly, as
we have already discussed, the $\ket{s(r,s,t)}^\pm$ vectors satisfy
the {\it massive\/} \hw{} conditions and, thus, conceal the fact that
the submodule generated from them is actually topological. Secondly,
these vectors may not even generate the entire of the topological
submodule generated from the respective~$\ket{E(r,s,t)}^\pm$. Indeed,
while in the general position we have
\begin{equation}
  \cG_{-r+1}\,\ldots\,\cG_0\,\ket{s(r,s,t)}^+=
  C^\pm(r,s,t)\,\ket{E(r,s,t)}^+
\end{equation}
(where $C$ is a scalar factor) and, thus, the respective singular
vectors \req{ST} and \req{Tplus},~\req{Tminus} generate the same
submodule, yet in degenerate cases the action of one of the $\cG_n$ in
the last formula may give the vanishing result, in which case
$\ket{s(r,s,t)}^\pm$ would generate only a submodule in the submodule
generated from $\ket{E(r,s,t)}^+$.  Therefore, working with the
singular vectors that satisfy \hw{} conditions \req{massivehw} makes
it necessary to introduce subsingular vectors in topological Verma
modules in order to completely describe the structure of submodules.
On the other hand, having the $\ket{E(r,s,t)}^\pm$ singular vectors at
our disposal makes the subsingular vector superfluous, namely an
artifact of having chosen $\ket{s(r,s,t)}^\pm$ to represent singular
vectors~\cite{[ST4]}.

\subsection{Singular vectors in massive Verma
  modules\label{subsec:MandCH}}\lvm The prescription to construct
singular vectors in a massive Verma module $\;\mU$ is to map the \hw{}
vector of the module into a generalized {\it topological\/} \hw{}
vector by means of the continued operators $g$ or $q$. In the
generalized topological Verma module, one then requires that a
(topological!)  singular vector exist. Further, one uses the $g$ or
$q$ operator to map this topological singular vector back into the
original module~$\;\mU$. This program is implemented as follows.

Given a massive \hw{} vector $\ket{h,\ell,t}$, let $\theta'$ and
$\theta''=-\theta'+ht-1$ be two roots of the equation
\begin{equation}
  \ell=-\theta h+\frac{1}{t}(\theta^2+\theta)\,.
  \label{ell}
\end{equation}
Then the states
\begin{equation}
  g(\theta',-1)\ket{h,\ell,t}\,,\qquad  q(-\theta',0)\ket{h,\ell,t}
  \label{1}
\end{equation}
and
\begin{equation}
  g(\theta'',-1)\ket{h,\ell,t}\,,\qquad q(-\theta'',0)\ket{h,\ell,t}\,,
  \label{2}
\end{equation}
are the generalized topological highest-weight vectors.  Whenever one
of the states~\req{1} (or~\req{2}) belongs to $\mU_{h,\ell,t}$, it
gives a charged singular vector in $\mU_{h,\ell,t}$ and we, thus,
recover Theorem~\ref{charged:thm}. Note that the condition for this to
be the case reproduces Eq.~\req{3rdconditions} for the parameters of
the \hw{} vector.  Other singular vectors in massive Verma modules can
be constructed using the following trick. We require one of the states
in~\req{1} or~\req{2} to admit a topological singular vector.
Using~\req{gmapextremal}--\req{htopnew}, we, thus, require that $h'$
or $h''$ take one of the values $\htop^\pm(r,s,t)$, see
Lemma~\ref{toplemma}. Therefore, $\theta'$ and $\theta''$ must be
equal to
\begin{equation}\new
  \begin{array}{rclcl}
    \theta_1(r, s, h,t)
    &=&\frac{t}{2}(h-\hminus(r,s,t))\,,\\
    \theta_2(r, s, h,t)
    &=&\frac{t}{2}(h-1-\hplus(r,s,t))\,.
  \end{array}
  \label{theta1theta2}
\end{equation}
Note that using these values of $\theta$ in~\req{ell}, we recover
Eq.~\req{theell}. Thus, along with \req{3rdconditions}, we have
recovered all of the zeros of the Ka\v c determinant~\cite{[BFK]}. The
charged singular vector are already given by the explicit
construction~\req{thirdE}, while the above derivation of~\req{theell}
suggests how the massive singular vectors, too, can be constructed
explicitly.

Let, for definiteness, the first of the states~\req{1} admit the
topological singular vector $\ket{E(r,s,t)}^-$, Eq.~\req{Tminus}. Then
$\theta'=\theta_1(r,s,h,t)$ and, as is easy to see, the second state
from~\req{1} admits a $\ket{E(r,s,t)}^+$ topological singular vector.
We thus obtain two singular vectors on the generalized topological
\hw{} states~\req{1}:
\begin{equation}\new
  \begin{array}{l}
    \cE^{-,\theta_1(r,s,h,t)}(r,s,t)\,g(\theta_1(r,s,h,t),-1)\,
    \ket{h,\theel(r,s,h,t),t}\,,\\
    \cE^{+,\theta_2(r,s,h,t)}(r,s+1,t)\,q(-\theta_2(r,s,h,t),0)\,
    \ket{h,\theel(r,s,h,t),t}
  \end{array}
  \label{predvectors}
\end{equation}
Mapping these singular vectors back to the original massive Verma
module $\mU_{h,\theel(r,s,h,t),t}$ by the appropriate $g$- or
$q$-operator, we obtain
\begin{thm}\mbox{}\nopagebreak
  \begin{enumerate}
    \addtolength{\parskip}{-6pt}
    
  \item For generic $h$ and $t$, representatives of the massive
    singular vector in $\mU_{h,\theel(r,s,h,t),t}$ read
    \begin{equation}\kern-8pt\new
      \begin{array}{rcl}
        \ket{S(r,s,h,t)}^-\kern-6pt&=&\kern-6pt
        g(-rs,r+\theta_1(r,s,h,t)-1)\,
        \cE^{-,\theta_1(r,s,h,t)}(r,s,t)\,g(\theta_1(r,s,h,t),-1)\,
        \ket{h,\theel(r,s,h,t),t}\,,
        \\
        \ket{S(r,s,h,t)}^+\kern-6pt&=&\kern-6pt
        q(1-rs,r-\theta_2(r,s,h,t)-1)
        \cE^{+,\theta_2(r,s,h,t)}(r,s,t)
        q(-\theta_2(r,s,h,t),0)
        \ket{h,\theel(r,s,h,t),t},
      \end{array}
      \label{Sgen}
    \end{equation}
    where $\theta_1(r, s, h, t)$ and $\theta_2(r, s, h, t)$ are given
    by~\req{theta1theta2} and $\cE^{\pm,\theta}(r,s,t)$ denotes the
    spectral flow transform \req{U} of the topological singular vector
    operator $\cE^{\pm}(r,s,t)$.
    
  \item The RHSs of \req{Sgen} evaluate as elements of $\;\mU_{h,
      \theel(r, s, h, t),t}$ and satisfy the twisted massive \hw{}
    conditions
    \begin{equation}\new
      \begin{array}{l}
        \cQ_{\geq1\mp rs}\,\ket{S(r,s,h,t)}^\pm=
        \cH_{\geq1}\,\ket{S(r,s,h,t)}^\pm=\cL_{\geq1}\,\ket{S(r,s,h,t)}^\pm=
        \cG_{\geq\pm rs}\,\ket{S(r,s,h,t)}^\pm=0\,,\\
        \cL_0\,\ket{S(r,s,h,t)}^\pm=
        \theel^\pm(r,s,h,t)\,\ket{S(r,s,h,t)}^\pm\,,\\
        \cH_{0}\,\ket{S(r,s,h,t)}^\pm=(h \mp r s)\,\ket{S(r,s,h,t)}^\pm
      \end{array}
      \label{hwsing}
    \end{equation}
    with
    \begin{equation}\new
      \begin{array}{rcl}
        \theel^-(r,s,h,t)\kern-4pt&=&\kern-4pt
        \theel(r,s,h,t) + \half (r s + 1) (r s + 2) - 1\,,\\
        \theel^+(r,s,h,t)\kern-4pt&=&\kern-4pt
        \theel(r,s,h,t) + \half r s (r s + 1)
      \end{array}
    \end{equation}
    In the generic case, either of the states $\ket{S(r,s,h,t)}^\pm$
    generates the entire massive Verma submodule; in particular, all
    of the dense $\cG/\cQ$-descend\-ants of~$\ket{S(r,s,h,t)}^+$ and
    $\ket{S(r,s,h,t)}^+$ are on the same extremal subdiagram (the
    extremal diagram of the submodule) and coincide up to numerical
    factors whenever they are in the same grade:
    \begin{equation}\new\kern-6pt
      \begin{array}{ll}
        c_-(i, h, t)\,\cQ_{i+1-rs}\ldots\cQ_{rs}\,\ket{S(r,s,h,t)}^-=
        c_+(i, h, t)\,\cG_{rs-i}\ldots\cG_{rs-1}\,\ket{S(r,s,h,t)}^+,
        \kern-4pt&\kern-4pt i=0,\ldots,2rs\,,\\
        c_-(i, h, t)\,\cG_{-rs+i}\ldots\cG_{-rs-1}\,\ket{S(r,s,h,t)}^-=
        c_+(i, h, t)\,\cG_{-rs+i}\ldots\cG_{rs-1}\,\ket{S(r,s,h,t)}^+,
        \kern-4pt&i\leq-1\,,\\
        c_-(i, h, t)\,\cQ_{rs-i}\ldots\cQ_{rs}\,\ket{S(r,s,h,t)}^-=
        c_+(i, h, t)\,\cQ_{rs-i}\ldots\cG_{-rs}\,\ket{S(r,s,h,t)}^+,
        \kern-4pt&i\geq2rs+1\,,
      \end{array}
      \label{compare}
    \end{equation}
    where the numerical coefficients $c_\pm(i,h,t)$ are ($r$- and
    $s$-dependent) polynomials in~$h$ and~$t$.
  \end{enumerate}
\end{thm}

Thus, whenever the $h$ and $t$ parameters of the massive Verma module
$\mU_{\theel(r,s,h,t),h,t}$ are in the general position, the maximal
submodule can be generated from $\ket{S(r,s,h,t)}^-$ as well as from
$\ket{S(r,s,h,t)}^+$. In the generic case, moreover, the top-level
representative of the extremal diagram of the massive submodule
generates that very same submodule as~$\ket{S(r,s,h,t)}^\pm$. In a
number of degenerate cases, however, the situation changes. The
complete classification of the degenerate cases will be given
in~\cite{[ST4]}, while here, in Sec.~\ref{sec:Algebra}, we consider
two basic cases of such a degeneration.

\section{Evaluating $\N2$ singular vectors in the Verma
  form\label{sec:Algebra}}\lvm In this section, we use the algebraic
properties of the continued operators from section
\ref{subsec:AlgRules} in order to formulate the recipe of rewriting
the general formulae for the massive and topological singular vectors
as polynomials in the usual Verma module creation operators acting on
the highest-weight state.

\subsection{The topological singular
  vectors\label{subsec:Top}}\lvm In this subsection, we show how
topological singular vectors \req{Tplus} and~\req{Tminus} can be
recast into the Verma form, i.e., into polynomials in the standard
creation operators $\cL_{\leq-1}$, $\cH_{\leq-1}$, $\cG_{\leq-1}$, and
$\cQ_{\leq-1}$ in the topological Verma module.

To this end, we formulate a recursive procedure which is based on
relations \req{recursion}.  We concentrate on the $\cE^+$ vector for
definiteness.  Assume that $\cE^-(r,s-1,t)$ does already have the
conventional Verma form (i.e., is a polynomial in creation operators).
Then, for any $\theta$ (which we will actually take to be $(s-1)t-r$),
the spectral-flow-transformed singular vector operator
$\cE^{-,\theta}(r,s-1,t)$ also rewrites as a polynomial in $\cL_{-m}$,
$\cH_{-m}$, $m\in\oN$, $\cG_{-m+\theta}$, and $\cQ_{-m-\theta}$, with
$m\in\oN$,

We now use gluing rules~\req{Glue} to rewrite the corresponding
formula from~\req{recursion} as
\begin{equation}
  \cE^+(r,s,t)=
  g(-r,(s-1)t-r-1)\,
  \cG_{(s-1)t-r}\ldots\cG_{(s-1)t-1}
  \,\cE^{-,(s-1)t-r}(r,s-1,t)\,g((s-1)t-r,-1)\,.
  \label{work99}
\end{equation}
Recall also that this singular vector operator is to be applied to the
\hw{} vector $\kettop{\hplus(r,s,t),t}$. Then, in accordance with
Lemma~\ref{lemma:HW}, we observe that each of the operators
$\cG_{(s-1)t-r}$, $\ldots$, $\cG_{(s-1)t-1}$ annihilates the {\it
  generalized topological \hw{} state\/}
$g((s-1)t-r,-1)\,\kettop{\hplus(r,s,t),t}$. We, thus, can commute all
of these operators in \req{work99} through the topological singular
vector operator $\cE^{-,(s-1)t-r}(r,s-1,t)$ and kill them as soon as
they reach the $g$-operator on the right. The topological singular
vector thus rewrites as
\begin{equation}
  \ket{E(r,s,t)}^+ = g(-r,(s-1)t-r-1)\,
  \cP(\cL,\cH,\cG,\cQ)\,g((s-1)t-r,-1)\,\kettop{\hplus(r,s,t),t}
  \label{work100}
\end{equation}
where $\cP$ is a polynomial in $\cL_{-m}$, $\cH_{-m}$,
$\cG_{-m+(s-1)t-r}$, and $\cQ_{-m-(s-1)t+r}$, with $m\in\oN$. Since
the topological singular vector $\ket{E(r,s,t)}^-$ has relative charge
$-r$, the result of commuting its singular vector operator with $r$
modes of $\cG$ is $\cH_0$-neutral; in other words, $\cP$ has zero
charge.

Observe further that any mode $\cG_\mu$ that can be encountered in
$\cP$ is necessarily such that formulae \req{leftkill} apply to the
product $g(-r,(s-1)t-r-1)\,\cG_\mu$. In this way, all of the
$\cG$-modes can be commuted to the left, after which they give the
vanishing contribution to~\req{work100}. Therefore, the singular
vector rewrites as
\begin{equation}
  \ket{E(r,s,t)}^+ = g(-r,(s-1)t-r-1)\,
  \bar\cP(\cL,\cH)\,g((s-1)t-r,-1)\,\kettop{\hplus(r,s,t),t}\,.
  \label{work101}
\end{equation}
In what follows, we refer to the thus obtained operator polynomial
$\cP(\cL,\cH)$ as the {\it skeleton\/} of the corresponding singular
vector. In~\req{work101}, had it not been for the skeleton in the
middle, we would have used the formula
$$\new
\begin{array}{rcl}
  g(-r,(s-1)t-r-1)\cdot g((s-1)t-r,-1)\,
  \kettop{\hplus(r,s,t),t}&=&g(-r,-1)\,\kettop{\hplus(r,s,t),t}\\
  {}&=&\prod_{i=-r}^{-1}\cG_i\,\kettop{\hplus(r,s,t),t}
\end{array}
$$
to glue two $g$-operators together, after which they reduce to the
Verma form. In fact, Eqs.~\req{LLeft} allow us to commute the $\cL$-
and $\cH$-modes on the left, after which the two $g$-operators meet
and one can use
$$
g(j,(s-1)t-r-1)\cdot g((s-1)t-r,-1)\,
\kettop{\hplus(r,s,t),t}=g(j,-1)\,\kettop{\hplus(r,s,t),t}
$$
Thus, the singular vector rewrites as
\begin{equation}
  \ket{E(r,s,t)}^+ =\left( \bar\cP(\cL,\cH) +
    \sum_{j=1}^{r(s-2)}
    \Bigl(\prod_{i=1}^{r+j}\cG_{\mu_i}\Bigr)
    \tilde \cP_j(\cL,\cH)\,g(j,-1)\right)\,
  \kettop{\hplus(r,s,t),t}\,,
  \label{weird}
\end{equation}
where $\cP_j$ are polynomials in $\cL_{-m}$, $\cH_{-m}$, and
$\cG_{-m}$, $m\in\oN$. To the terms that involve the $g$-operators of
a negative integral length, we now apply
Eqs.~\req{gtrans}--\req{Lambdaboth}.

Starting with higher $j$, we, thus, replace
\begin{equation}
  g(j, -1)\,\kettop{\htop^+(r,s,t),t}= \,
  {t\over2(j - 1)(\htop^+(r,s,t) t - j)}\,
  \cQ_{1 - j}\,g(j - 1, -1)\,
  \kettop{\htop^+(r,s,t),t}\,,\quad j\geq2
  \label{Fill1}
\end{equation}
and commute the product $(\prod\cG)$ in \req{weird} to the right.
Then, in some terms the combination $(\prod\cG)\,g(j-1, -1)$ would
allow us to apply \req{Glue}, which would give one of the $g(j-2,-1)$,
$g(j-3,-1)$, $\ldots$, $g(0,-1)$ operators.  In the latter case
($g(0,-1)=1$) the $g$-operator will have disappeared. To the terms
that would still contain $g(j-1, -1)$, $\ldots$, $g(2, -1)$, we apply
\req{Fill1} and the corresponding rearrangements again, until we end
up with having, on top of a state from the Verma module
$\mV_{\hplus(r,s,t),t}$, only the terms that contain $g(1, -1)$.
However, {\it all the latter cancel against the different terms\/},
and we are therefore left with a state in $\mV_{\hplus(r,s,t),t}$.

This vanishing property is a non-trivial feature of the whole scheme;
a related fact is that applying formula~\req{gtrans} to the evaluation
of topological singular vectors does always preserve the coefficients
in front of the different terms in the ring $t^{-M}\cdot\oQ[t]$
(polynomials with rational coefficients times a certain negative power
of $t$ that may come out of the highest-weights \req{hplushminus}),
i.e.\ no rational dependence on $t$ arises, apart from a possible
$t^{-M}$.

Thus, having assumed that the inner singular vector operators in
\req{work99} are already in the Verma form, we see that the whole
expression \req{work99} is in turn evaluated in the Verma form.  The
argument now applies recursively, until we reach the respective
simplest singular vector
\begin{equation}\new
  \begin{array}{rclccrcl}
    \ket{E(r,1,t)}^+&=&\cG_{-r}\,\ldots\,\cG_{-1}\,
    \ket{\hplus(r,1,t),t}_{\rm top}\,,&&
    \cE^+(r,1,t)&=&\cG_{-r}\,\ldots\,\cG_{-1}\,\\
    \ket{E(r,1,t)}^-&=&\cQ_{-r}\,\ldots\,\cQ_{-1}\,
    \ket{\hminus(r,1,t),t}_{\rm top}\,,&&
    \cE^-(r,1,t)&=&\cQ_{-r}\,\ldots\,\cQ_{-1}
  \end{array}
  \label{s1plus}
\end{equation}

Evaluation in the $\cE^-$-case is completely similar.

\subsection{The massive singular vectors\label{subsec:Massive}}\lvm As
follows from \req{Sgen}, the representatives $\ket{S^\pm(r,s,h,t)}$ of
a massive singular vector can be derived from the appropriate
topological singular vectors $\ket{E^\pm(r,s,t)}$. As before, an
important point is that, at a certain stage in the evaluation, all
that remains of the topological singular vector operator is the
skeleton -- a polynomial in modes of {\it only\/} $\cL_m$ and~$\cH_m$.

Consider for definiteness how the massive singular vectors are
evaluated as Verma module elements using the form $\ket{S(r,s,h,t)}^-$
from Eq.~\req{Sgen}.  First of all, we rewrite
$$
g(-rs,r+\theta_1(r,s,h,t)-1)\,
\cE^{-,\theta_1(r,s,h,t)}(r,s,t)\,g(\theta_1(r,s,h,t),-1)
$$
as
$$
g(-rs,\theta_1(r,s,h,t)-1)\,\cG_{\theta_1(r,s,h,t)}\ldots
\cG_{\theta_1(r,s,h,t)+r-1}\,
\cE^{-,\theta_1(r,s,h,t)}(r,s,t)\,g(\theta_1(r,s,h,t),-1)\,.
$$
We now observe that the modes $\cG_{\theta_1(r,s,h,t)}$, \ldots,
$\cG_{\theta_1(r,s,h,t)+r-1}$ annihilate the state
$\,g(\theta_1(r,s,h,t),-1)\cdot \ket{h,\theel(r,s,h,t),t}$. We, thus,
commute all these modes to the right in every operator monomial. Then,
the remaining $\cG$-modes are annihilated by
$g(-rs,r+\theta_1(r,s,h,t)-1)\,$ in accordance with~\req{leftkill}.
After commuting the $\cG$-modes to the left and then dropping the
$\cG$-dependent monomials, the operator $g(0,\theta_1(r,s,h,t)-1)$ is
separated from the right one, $\,g(\theta_1(r,s,h,t),-1)$, by modes of
only $\cL$ and $\cH$ (the {\it skeleton\/}).  We now apply
Eqs.~\req{LLeft} repeatedly.  Then the two $g$-operators meet and
produce, in accordance with \req{Glue} and~\req{integrallength},
either the product of $\cG$ modes or the identity operator.  Thus we
are left with a singular vector in the Verma
module~$\,\mU_{h,\theel(r,s,h,t),t}$.

\smallskip

The $+$-representative of the massive singular vector is evaluated in
a similar way starting with the topological singular vector
$\ket{S(r,s,t)}^+$. Here, one first commutes $r$ modes
$\cQ_{-\theta_2(r,s,h,t)}$, \ldots, $\cQ_{r-\theta_2(r,s,h,t)-1}$ to
the right until they annihilate the state $q(-\theta_2(r,s,h,t),0)\,
\ket{h,\theel(r,s,h,t),t}$. Then all the remaining $q$-modes are
commuted to the left, where, again, they are annihilated by
$q(1-rs,-\theta_2(r,s,h,t)-1)$.

In this way, we obtain massive singular vectors as elements of the
corresponding Verma module.

\subsection{Evaluating massive singular vectors: an example
  \label{subsec:Examples}}\lvm
Here, we give an example of the evaluation of massive singular
vectors.  As the length of the expressions grows rapidly with $r$ and
$s$, we restrict ourselves to level-3 singular vector
$\ket{S(1,3,h,t)}^+$ and $\ket{S(1,3,h,t)}^-$. We follow the strategy
of section \ref{subsec:MandCH}: deriving the massive singular vector
from the corresponding topological one. Then, we demonstrate that
$\ket{S(1,3,h,t)}^+$ and $\ket{S(1,3,h,t)}^-$ generate the same
submodule.

Consider first $\ket{S(1,3,h,t)}^-$.  The starting point is the
topological singular vector $\ket{E(1,3,t)}^-$ (see \req{Tminus})
written in the polynomial form (see~\cite{[ST2]} for the examples of
calculations of topological singular vectors):
\begin{equation}\new
  \begin{array}{l}
    \ket{E(1,3,t)}^-=\Bigl((4 t + 16 t^2) \cQ_{-3}  +
    (-4 + 8 t^2) \cH_{-2}    \cQ_{-1}  +
    (-12 t + 16 t^2) \cH_{-1}    \cQ_{-2} -4 t \cL_{-2} \cQ_{-1}\\
    -12 t \cL_{-1}    \cQ_{-2}  +
    (4 - 12 t + 8 t^2) \cH_{-1}^2      \cQ_{-1}  +
    (8 - 12 t) \cL_{-1}    \cH_{-1}    \cQ_{-1}  +
    4 \cL_{-1}^2     \cQ_{-1}\Bigr)\,\kettop{-3 + \frac{2}{t}, t}
  \end{array}
\end{equation}
Now, in accordance with the procedure described in
Sec.~\ref{subsec:Massive}, we calculate the skeleton
of~$\ket{E(r,s,t)}^-$.  To this end, we write the massive singular
vector to be evaluated as
\begin{eqnarray}
  \ket{S(1,3,h,t)}^-&=&
  g(-3,\frac{t}{2}(h-\hminus(1,3,t))-1)\cdot{}\\
  {}&{}&\qquad
  \underbrace{\cG_{\frac{t}{2}(h-\hminus(1,3,t))}\,
    \cE^{-,\frac{t}{2}(h-\hminus(1,3,t))}}\cdot
  g(\frac{t}{2}(h-\hminus(1,3,t)),-1)\,
  \ket{h,\theel(1,3,h,t),t}\,,\nonumber
\end{eqnarray}
To evaluate the underbraced factors, where
$\cG_{\frac{t}{2}(h-\hminus(1,3,t))}$ has to be commuted on the right
(after which it vanishes), we can consider the
expression~$\cG_1\,\ket{E(r,s,t)}^-$ before the spectral flow
transform and rearrange each monomial in such a way that every mode
of~$\cG$ is moved on the right of every mode of~$\cQ$; then, dropping
all the monomials that contain~$\cG$ or~$\cQ$ modes, we evaluate the
underbraced factors as
\begin{equation}\new
  \begin{array}{l}
    (16 + 48 t + 32 t^2) \cH_{-3}
    + (-16 t + 32 t^2) \cL_{-3}
    + (-24 + 48 t^2) \cH_{-2}  \cH_{-1}
    + (-32 t + 32 t^2) \cL_{-2}    \cH_{-1}\\
    - 32 t \cL_{-2}    \cL_{-1}
    + (-24 - 24 t + 16 t^2) \cL_{-1}    \cH_{-2}
    + (8 - 24 t + 16 t^2) \cH_{-1}^3
    + (24 - 48 t + 16 t^2) \cL_{-1}    \cH_{-1}^2 \\
    + (24 - 24 t) \cL_{-1}^2  \cH_{-1}
    + 8 \cL_{-1}^3
  \end{array}
\end{equation}
To obtain the skeleton $\cP(\cL,\cH)$, this must be subjected to the
spectral flow transform with the spectral
parameter~$\frac{t}{2}(h-\hminus(1,3,t))$.\footnote{ Note that the
  order of operations is irrelevant; one can subject the
  vector~$\ket{E(r,s,t)}^-$ to the spectral flow transform and then
  calculate the skeleton or calculate the skeleton
  of~$\ket{E(r,s,t)}^-$ and then subject it to the spectral flow with
  the same spectral parameter; the result is the same in both cases.}
This gives
\begin{eqnarray}
  \ket{S(1,3,h,t)}^-&=&
  g(-3,\frac{t}{2}(h-\hminus(1,3,t))-1)\cdot{}
  \Bigl((-16 t + 32 t^2)
  \cL_{-3} - 32 t
  \cL_{-2} \cL_{-1} + {}\nonumber\\{}&{}&{}
  (24 t + 8 h t + 72 t^2 + 24 h t^2 + 48 t^3 + 16 h t^3)
  \cH_{-3}
  + (-3 t^3 - h t^3 + 3 h^2 t^3 + h^3 t^3)
  \cH_{-1}^3 - {}\nonumber\\{}&{}&{}
  (6 t^2 + 20 h t^2 + 6 h^2 t^2 + 24 h t^3 + 8 h^2 t^3)
  \cH_{-2}    \cH_{-1}
  - (16 t^2 + 16 h t^2)
  \cL_{-2}    \cH_{-1} - {}
  \nonumber\\{}&{}&{}
  (28 t + 12 h t + 32 t^2 + 16 h t^2)
  \cL_{-1}    \cH_{-2}
  - (2 t^2 - 12 h t^2 - 6 h^2 t^2)
  \cL_{-1}  \cH_{-1}^2 + {}
  \label{gskelg}\\{}&{}&{}
  (12 t + 12 h t)
  \cL_{-1}^2    \cH_{-1}
  + 8 \cL_{-1}^3 \Bigr)\cdot
  g(\frac{t}{2}(h-\hminus(1,3,t)),-1)\ket{h,\theel(1,3,h,t),t}
  \nonumber
\end{eqnarray}
Using formulae~\req{LLeft} and then~\req{Glue} and
\req{integrallength}, we evaluate~\req{gskelg} in the polynomial form
\begin{eqnarray}
  &&\ket{S(1,3,h,t)}^-=\Bigl(
  - (6 + 5 t + h t) (32 + 20 t + 12 h t + 9 t^2 + 6 h t^2 + h^2 t^2)
  \cG_{-6}\cG_{-2}\cG_{-1}\nonumber\\ &&
  + 2 (-48 - 44 t - 44 h t + 16 t^2 - 20 h t^2 - 12 h^2 t^2 + 27 t^3 + 9 h t^3 -
  3 h^2 t^3 - h^3 t^3)
  \cG_{-5}\cG_{-3}\cG_{-1} \nonumber\\ &&
  + (3 + h) t (-2 + 3 t - h t) (4 - 5 t + h t)
  \cG_{-4}    \cG_{-3}    \cG_{-2}
  + 8 (3 + h) t (1 + t) (1 + 2 t)
  \cH_{-3}    \cG_{-3}    \cG_{-2}    \cG_{-1}\nonumber\\ &&
  + 2 t (28 + 12 h + 35 t + 26 h t + 3 h^2 t + 12 h t^2 + 4 h^2 t^2)
  \cH_{-2}    \cG_{-4}    \cG_{-2}    \cG_{-1}\nonumber\\ &&
  + t (72 + 72 h + 14 t + 84 h t + 30 h^2 t - 9 t^2 + 21 h t^2 +
  17 h^2 t^2 + 3 h^3 t^2)
  \cH_{-1}    \cG_{-5}    \cG_{-2}    \cG_{-1}\nonumber\\ &&
  + t (24 + 24 h - 14 t + 28 h t + 18 h^2 t - 9 t^2 - 27 h t^2 +
  h^2 t^2 + 3 h^3 t^2)
  \cH_{-1}    \cG_{-4}    \cG_{-3}    \cG_{-1}\nonumber\\ &&
  + 2 (72 + 54 t + 30 h t + 15 t^2 + 14 h t^2 + 3 h^2 t^2)
  \cL_{-1}    \cG_{-5}    \cG_{-2}    \cG_{-1}
  + 16 t (4 + t + h t)
  \cL_{-2}    \cG_{-4}    \cG_{-2}    \cG_{-1}\nonumber\\ &&
  + 2 (24 + 10 t + 18 h t - 17 t^2 - 2 h t^2 + 3 h^2 t^2)
  \cL_{-1}    \cG_{-4}    \cG_{-3}    \cG_{-1}
  + 16 t (-1 + 2 t)
  \cL_{-3}    \cG_{-3}    \cG_{-2}    \cG_{-1}\nonumber\\ &&
  - 2 (3 + h) t^2 (1 + 3 h + 4 h t)
  \cH_{-2}    \cH_{-1}    \cG_{-3}    \cG_{-2}    \cG_{-1}
  - 32 t
  \cL_{-2}    \cL_{-1}    \cG_{-3}    \cG_{-2}    \cG_{-1}\nonumber\\ &&
  + t^2 (4 - 24 h - 12 h^2 + 9 t + 3 h t - 9 h^2 t - 3 h^3 t)
  \cH_{-1}^2  \cG_{-4}    \cG_{-2}    \cG_{-1}
  - 16 (1 + h) t^2
  \cL_{-2}    \cH_{-1}    \cG_{-3} \cG_{-2} \cG_{-1} \nonumber\\ &&
  - 4 t (7 + 3 h + 8 t + 4 h t)
  \cL_{-1}    \cH_{-2}    \cG_{-3}    \cG_{-2}    \cG_{-1}
  + 4 t (-12 - 12 h + t - 6 h t - 3 h^2 t)
  \cL_{-1}    \cH_{-1}    \cG_{-4}    \cG_{-2}    \cG_{-1}  \nonumber\\ &&
  - 12 (4 + t + h t)
  \cL_{-1}^2  \cG_{-4}    \cG_{-2}    \cG_{-1}
  +  (-1 + h) (1 + h) (3 + h) t^3
  \cH_{-1}^3  \cG_{-3}    \cG_{-2}    \cG_{-1}               \nonumber\\ &&
  + 2 (-1 + 6 h + 3 h^2) t^2
  \cL_{-1}    \cH_{-1}^2  \cG_{-3}    \cG_{-2}    \cG_{-1}
  + 12 (1 + h) t
  \cL_{-1}^2  \cH_{-1}    \cG_{-3}    \cG_{-2}    \cG_{-1} \nonumber\\ &&
  + 8
  \cL_{-1}^3  \cG_{-3}    \cG_{-2}    \cG_{-1}\Bigr)
  \ket{h,\theel(1,3,h,t),t}
\end{eqnarray}

In the same way, the vector $\ket{S(1,3,h,t)}^+$ can be written in the
polynomial form
\begin{eqnarray}
  &&\ket{S(1,3,h,t)}^+=\Bigl(
  (-6 - 5 t + h t) (32 + 20 t - 12 h t + 9 t^2 - 6 h t^2 + h^2 t^2)
  \cQ_{-5}    \cQ_{-1}    \cQ_{0}\nonumber\\ &&
  + 2(-48 - 44 t + 44 h t + 16 t^2 + 20 h t^2 - 12 h^2 t^2 + 27 t^3 - 9 h t^3 -
  3 h^2 t^3 + h^3 t^3)
  \cQ_{-4}    \cQ_{-2}    \cQ_{0} \nonumber\\ &&
  + (-3 + h) t (-2 + 3 t + h t) (-4 + 5 t + h t)
  \cQ_{-3}    \cQ_{-2}    \cQ_{-1}
  + 8 t (1 + 2 t) (1 + h - 3 t + h t)
  \cH_{-3}    \cQ_{-2}    \cQ_{-1}    \cQ_{0}      \nonumber\\ &&
  + 2 t (4 + 12 h - 27 t + 18 h t - 3 h^2 t + 12 h t^2 - 4 h^2 t^2)
  \cH_{-2}    \cQ_{-3}    \cQ_{-1}    \cQ_{0}        \nonumber\\ &&
  + t (-72 + 72 h - 14 t + 84 h t - 30 h^2 t + 9 t^2 + 21 h t^2 -
  17 h^2 t^2 + 3 h^3 t^2)
  \cH_{-1}    \cQ_{-4}    \cQ_{-1}    \cQ_{0} \nonumber\\ &&
  + t (-24 + 24 h + 14 t + 28 h t - 18 h^2 t + 9 t^2 - 27 h t^2 -
  h^2 t^2 + 3 h^3 t^2)
  \cH_{-1}    \cQ_{-3}    \cQ_{-2}    \cQ_{0}   \nonumber\\ &&
  + 2 (72 + 54 t - 30 h t + 15 t^2 - 14 h t^2 + 3 h^2 t^2)
  \cL_{-1}    \cQ_{-4}    \cQ_{-1}    \cQ_{0}
  + 16 t (-1 + 2 t)
  \cL_{-3}    \cQ_{-2}    \cQ_{-1}    \cQ_{0}\nonumber\\ &&
  + 2 (24 + 10 t - 18 h t - 17 t^2 + 2 h t^2 + 3 h^2 t^2)
  \cL_{-1}    \cQ_{-3}    \cQ_{-2}    \cQ_{0}
  + 16 t (4 + t - h t)
  \cL_{-2}    \cQ_{-3}    \cQ_{-1}    \cQ_{0}\nonumber\\ &&
  + 2 t^2 (5 + 2 h - 3 h^2 + 12 h t - 4 h^2 t)
  \cH_{-2}    \cH_{-1}    \cQ_{-2}    \cQ_{-1}    \cQ_{0}
  + 16 (1 - h) t^2 \cL_{-2}
  \cH_{-1}    \cQ_{-2}    \cQ_{-1}    \cQ_{0} \nonumber\\ &&
  + t^2 (4 + 24 h - 12 h^2 + 9 t - 3 h t - 9 h^2 t + 3 h^3 t)
  \cH_{-1}^2  \cQ_{-3}    \cQ_{-1}    \cQ_{0}
  - 32 t \cL_{-2}
  \cL_{-1}    \cQ_{-2}    \cQ_{-1}    \cQ_{0}     \nonumber\\ &&
  + 4 t (-1 - 3 h + 8 t - 4 h t)
  \cL_{-1}    \cH_{-2}    \cQ_{-2}    \cQ_{-1}    \cQ_{0}
  + 4 t (12 - 12 h - t - 6 h t + 3 h^2 t)
  \cL_{-1}    \cH_{-1}    \cQ_{-3} \cQ_{-1} \cQ_{0}     \nonumber\\ &&
  + 12 (-4 - t + h t)
  \cL_{-1}^2  \cQ_{-3}    \cQ_{-1}    \cQ_{0}
  + (-3 + h) (-1 + h) (1 + h) t^3
  \cH_{-1}^3  \cQ_{-2}    \cQ_{-1}    \cQ_{0}   \nonumber\\ &&
  + 2 (-1 - 6 h + 3 h^2) t^2
  \cL_{-1}    \cH_{-1}    \cH_{-1}    \cQ_{-2}    \cQ_{-1}    \cQ_{0}
  + 12 (-1 + h) t
  \cL_{-1}^2  \cH_{-1}    \cQ_{-2}    \cQ_{-1}  \cQ_{0}\nonumber \\ &&
  + 8
  \cL_{-1}^3  \cQ_{-2}    \cQ_{-1}    \cQ_{0} \Bigr)\,
  \ket{h,\theel(1,3,h,t),t}
\end{eqnarray}

We now illustrate \req{compare} by an explicit calculation:
\begin{equation}\new
  \begin{array}{l}
    \cQ_{-3}\cQ_{-2}\cQ_{-1}\cQ_0\cQ_1\cQ_2\cQ_3\ket{S(1,3,h,t)}^-=\\
    \qquad\frac{1}{8 t^2}
    (3 + h)(-2 + 3 t - h t)(4 - 3 t + h t) (2 + 3 t + h t) (4 + 3 t + h t)
    (6 + 3 t + h t)\ket{S(1,3,h,t)}^+
  \end{array}
\end{equation}

Whenever one of the factors on the RHS vanishes, there is a charged
singular vector simultaneously with the massive one, and
$\ket{S(1,3,h,t)}^-$ is inside the twisted topological Verma submodule
generated from the charged singular vector. In particular,
$\ket{S(1,3,h,t)}^-$ does not generate the {\it massive\/} Verma
submodule then.  For example, in the case where $h=-3$, we have
\begin{equation}
  \cQ_{-1}\cQ_0\cQ_1\cQ_2\cQ_3\ket{S(1,3,-3,t)}^-=0
\end{equation}
whereas $\cQ_0\cQ_1\cQ_2\cQ_3\ket{S(1,3,-3,t)}^-\neq0$.  This means
that the state $\cQ_0\cQ_1\cQ_2\cQ_3\ket{S(1,3,-3,t)}^-$ is the
(twisted) topological \hw{} state and $\ket{S(1,3,-3,t)}^-\neq0$
generates the (twisted) topological Verma submodule rather than
massive one.  However, the entire massive Verma submodule {\it is\/}
generated from~$\ket{S(1,3,-3,t)}^+$. As regards
$\cQ_0\cQ_1\cQ_2\cQ_3\ket{S(1,3,-3,t)}^-$, we can apply to this state
the operator $g(2,0)$ of length $-1$. The state
$g(2,0)\,\cQ_0\cQ_1\cQ_2\cQ_3\ket{S(1,3,-3,t)}^-$ is to be evaluated
in accordance with the rules of Sec.~\ref{subsec:AlgRules}. In our
example, we then continue acting with the corresponding modes of
$\cQ$, and eventually recover the vector $\ket{S(1,3,-3,t)}^+$:
\begin{equation}
  \cQ_{-3}\cQ_{-2}g(2,0)\,\cQ_0\cQ_1\cQ_2\cQ_3\ket{S(1,3,-3,t)}^-=
  {24 (1 - 3 t) (-2 + 3 t)\over t^2 (1 + 3 t)}
  \ket{S(1,3,-3,t)}^+
\end{equation}
The case where $t=-\frac{1}{3}$ is the one of a yet higher
codimension, where there is yet another charged singular vector in the
module, and the structure of submodules is more complicated, as
described in~\cite{[ST4]}.

An even simpler example is provided by the massive singular vector
$\ket{S(1, 1, h, t)}^-$.  In this case, the topological singular
vector from which we start reads
\begin{equation}
  \ket{E(1, 1 ,t)}^- =
  \cQ_{-1}\,\kettop{-1 + \frac{2}{t}, t}
\end{equation}
The skeleton is simply $2\cL_{-1}+2\cH_{-1}$, and one easily obtains
\begin{equation}
  \ket{S(1, 1, h, t)}^- =
  \Bigl(-(1 + h)t\cG_{-2} +
  (1 + h)t\cH_{-1}  \cG_{-1} +
  2\cL_{-1}  \cG_{-1} \Bigr)
  \ket{h, -\half + \frac{h}{2} + \frac{t}{4} - \frac{1}{4}h^2t, t}
\end{equation}
One can also obtain, in a similar way,
\begin{equation}
  \ket{S(1, 1, h, t)}^+ =
  \Bigl( (-1 + h)t\cQ_{-1}  +
  (-1 + h)t\cH_{-1}  \cQ_{0}  +
  2\cL_{-1}  \cQ_{0}\Bigr)
  \ket{h, -\half + \frac{h}{2} + \frac{t}{4} - \frac{1}{4}h^2t, t}
\end{equation}
In order to ``compare'' $\ket{S(1, 1, h, t)}^-$ with $\ket{S(1, 1, h,
  t)}^+$ we evaluate $\cQ_1\,\cQ_0\,\ket{S(1, 1, h, t)}^-$.  We find
\begin{equation}\kern-6pt\new
  \begin{array}{rcl}
    \cQ_0\,\ket{S(1, 1, h, t)}^- \kern-6pt&=&\kern-6pt
    \Bigl(\half(1 - h)(1 + h)t(2 + t + ht)\cH_{-1}  +
    (1 - h)(2 + t + ht)\cL_{-1}  -
    (2 + t + ht)\cG_{-1}  \cQ_{0}\Bigr)\cdot{}\\
    {}\kern-6pt&{}&\kern-6pt
    \quad\ket{h, -\half + \frac{h}{2} + \frac{t}{4} - \frac{1}{4}h^2t, t}
  \end{array}
\end{equation}
and finally,
\begin{equation}
  \cQ_1\,\cQ_0\,\ket{S(1, 1, h, t)}^-
  =-\half(1 + h)(2 + t + ht) \ket{S(1,1,h,t)}^+.
\end{equation}

\subsection{Examples of singular vectors in codimension
  $\geq2$\label{subsec:Intersec}}\lvm For the complete classification
of degenerate cases, the reader is referred to~\cite{[ST4]}; here, we
give most characteristic examples, which would at the same time
illustrate how our general construction for singular vectors works.
We consider some simplest cases where charged and massive singular
vectors coexists in a massive Verma module and a simple case where two
singular vectors lie in the same grade.

\paragraph{Coexistence of a massive and a charged singular vector}
Let us consider the case where a charged singular vector
$\ket{E(n,h,t)}_{\rm ch}$ exists in $\mU_{h,\ell,t}$ simultaneously
with the massive singular vector labelled by two positive integers~$r$
and~$s=1$ (we choose $s=1$ for simplicity, in particular to simplify
the diagrams with which we illustrate this case).  This happens when
the \hw\ parameters of the massive Verma module $\mU_{h,\ell,t}$ are
$\ell=\ellmc(r,1,h,t)$ and $h=\hmc(r,1,h,t)$, with
\begin{equation}\new
  \begin{array}{l}
    \ellmc(r,1,h,t)=n(-\frac{n}{t}+\frac{r}{t}-1)\,,\\
    \hmc(r,1,h,t)=-\frac{2n-1}{t}+\frac{r}{t}-1\,,
  \end{array}
  \quad n\in\oZ\,,\quad r\in\oN\,.
\end{equation}
In this case the module $\mU_{h,\ell,t}$ contains a massive Verma
submodule~$\,\mU'$ and a submodule $\mC$ generated from the charged
singular vector (with $n$ chosen to be positive, for definiteness)
\begin{equation}
  \ket{E_{\rm ch}(n,\hmc(r,1,h,t),t)}=
  \cG_{-n}\,\ldots\,\cG_{-1}\,\ket{\hmc(r,1,h,t),\ellmc(r,1,h,t),t}\,.
\end{equation}
$\mC$ is the twisted topological Verma module
$\smV_{\hminus(r,1,t),t;-n}$.  Also, we have
$\mC'=\,\mU'\cap\mC=\smV_{\frac{1-r}{t}-1,t;r-n}$, which is generated
from the state
\begin{equation}
  \ket{T}=\cQ_{n-r}\ldots\cQ_{n-1}\,\cG_{-n}\ldots\cG_{-1}\,
  \ket{\hmc(r,1,h,t),\ellmc(r,1,h,t),t}\,.
  \label{Top}
\end{equation}

Now, states~\req{Sgen} take the form
\begin{equation}\kern-8pt\new
  \begin{array}{rcl}
    \ket{S(r,1,\hmc(r,1,h,t),t)}^-\kern-6pt&=&\kern-6pt
    g(-r,r-n-1)\,
    \cQ_{n-r}\ldots\cQ_{n-1}\,\cG_{-n}\ldots\cG_{-1}\,
    \ket{\hmc(r,1,h,t),\ellmc(r,1,h,t),t}\,,
    \\
    \ket{S(r,1,\hmc(r,1,h,t),t)}^+\kern-6pt&=&\kern-6pt
    q(1-r,n + t-1)\,
    \cG_{-n - t}\ldots\cG_{-n + r - t}\,q(n - r + t,0)\cdot{}\\
    {}&{}&\hfill
    \ket{\hmc(r,1,h,t),\ellmc(r,1,h,t),t}\,,
  \end{array}
\end{equation}
whence we see that the $\ket{S(r,1,\hmc(r,1,h,t),t)}^+$ representative
--- which is evaluated as an element of the module $\mU_{h,\ell,t}$ as
described in Sec.~\ref{subsec:Massive} --- generates the massive
submodule~$\,\mU'$.  On the other hand, whenever $2r\geq n$, the
representative $\ket{S(r,1,\hmc(r,1,h,t),t)}^-$ does not, because
$\ket{S(r,1,\hmc(r,1,h,t),t)}^-$ is a descendant of the
state~\req{Top}:
\begin{equation}\new
  \begin{array}{rcl}
    \ket{S(r,1,\hmc(r,1,h,t),t)}^-&=&
    \cG_{-r}\ldots \cG_{r-n-1}\,
    \cQ_{n-r}\ldots\cQ_{n-1}\,\cG_{-n}\ldots\cG_{-1}\,
    \ket{\hmc(r,1,h,t),\ellmc(r,1,h,t),t}\\
    {}&=&\cG_{-r}\ldots \cG_{r-n-1}\,\ket{T}\,.
  \end{array}
\end{equation}
Thus, whenever $2r\geq n$, the vector $\ket{S(r,1,\hmc(r,1,h,t),t)}^-$
generates the submodule~$\,\mC'$ rather than~$\,\mU'$.  The maximal
submodule in this case is $\mU'\cup\mC$; it is generated by two
singular vectors $\ket{E_{\rm ch}(n,\hmc(r,1,h,t),t)}$ and
$\ket{S(r,1,\hmc(r,1,h,t),t)}^+$.

The same mechanism leads to the failure of the top-level
representative of the singular vectors to generate the maximal
submodule whenever $r\geq n$.  In this case, the top-level
representative of the massive singular vector labelled by $(r,1)$ is
the descendant of the state~$\ket{T}$:
\begin{equation}
  \ket{s}=
  \cG_{0}\ldots \cG_{r-n-1}\,
  \cQ_{n-r}\ldots\cQ_{n-1}\,\cG_{-n}\ldots\cG_{-1}\,
  \ket{\hmc(r,1,h,t),\ellmc(r,1,h,t),t}
  = \cG_{0}\ldots \cG_{r-n-1}\,\ket{T}\,.
\end{equation}
and therefore belongs to the module $\mC'$.  Now, it is obvious that
the states of the form
\begin{equation}
  \cQ_{-r-m}\ldots\cQ_{-r}\ket{S(r,1,\hmc,t)}^+\qquad m\geq0
\end{equation}
cannot be generated from $\ket{s}$.  The top-level representative
generates the submodule~$\mC$ only.  Thus, in the conventional
approach, where one uses only the top-level representatives of the
extremal diagrams as `singular vectors', one has to use a subsingular
vector in order to generate the entire submodule $\mU'$.  Of course,
the subsingular vector is a descendant of the
state~$\ket{S(r,1,\hmc,t)}^+$ since the latter generates the
submodule~$\mU'$.  One can check that the state
\begin{equation}\new
  \begin{array}{rcl}
    \ket{\rm Sub}&=&
    \cG_{0}\ldots \cG_{r-n-1}\,\cG_{r-n+1}\ldots \cG_{r-1}\,
    \ket{S(r,1,\hmc(r,1,h,t),t)}^+\\
    {}&=&\cG_{0}\ldots \cG_{r-n-1}\,g(r-n+1,r-n-1)\,\ket{T}
  \end{array}
\end{equation}
is the subsingular vector.\footnote{The vector $\ket{\rm Sub}$ exists
  as an element of the massive Verma module because
  $\ket{S(r,1,\hmc(r,1,h,t),t)}^+$ is an element of the Verma module
  as described in Sec.~\ref{subsec:Massive}.  However, one can be
  interested in how the expression $g(r-n+1,r-n-1)\,\ket{T}$ can be
  rewritten as an element of the Verma module. To do this, one
  represents the expression
  $$
  g(r-n+1,r-n-1)\,\cQ_{n-r}\ldots\cQ_{n-1}\,\cG_{-n}\ldots\cG_{-1}\,
  \ket{\hmc(r,1,h,t),\ellmc(r,1,h,t),t}
  $$
  in the form
  $$
  g(r-n+1,-n-1)\,\cG_{-n}\ldots\cG_{r-n-1}\,
  \cQ_{n-r}\ldots\cQ_{n-1}\,g(-n,-1)\,
  \ket{\hmc(r,1,h,t),\ellmc(r,1,h,t),t}\,,
  $$
  to which one applies the procedure described in
  Secs.~\ref{subsec:Top} and~\ref{subsec:Massive}.}  Indeed,
\begin{equation}
  \cH_1\ket{\rm Sub}=\cG_{0}\ldots \cG_{r-n-2}\,\ket{T}=
  (-1)^{r-n-1}\,\frac{t}{2(r + t-1)}\,\cQ_{n-r+1}\,\ket{s}
\end{equation}
and other conditions can be checked by similar direct calculations.
Thus, $\ket{\rm Sub}$ satisfies the conventional annihilation
conditions~\req{massivehw} as soon as $\ket{s}$ is set to zero.

This can be illustrated in the following extremal diagram:
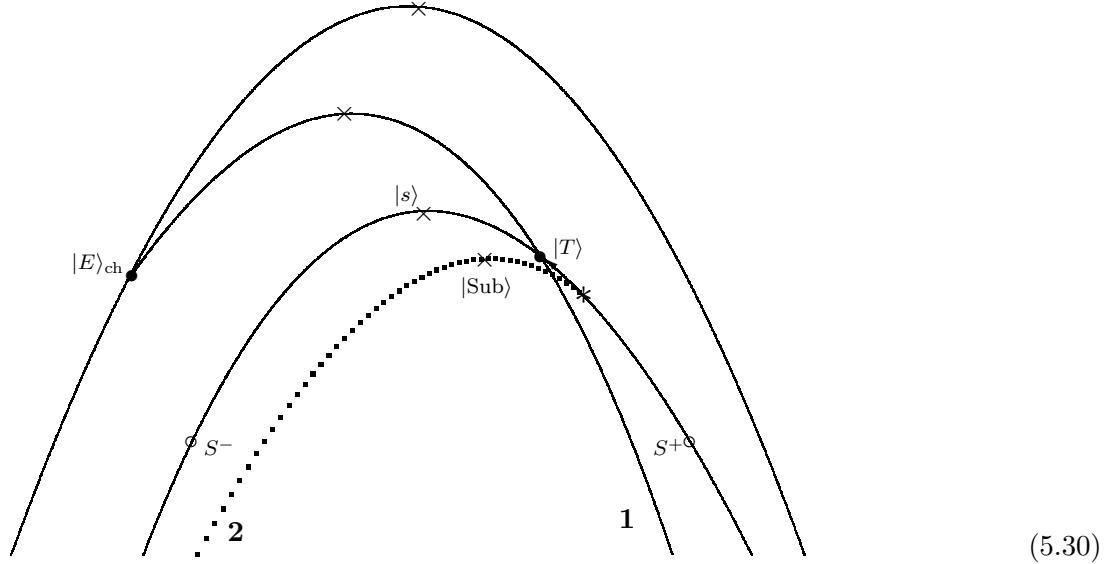
\begin{equation}
  \message{please wait...}
  \unitlength=1pt
  \begin{picture}(400,210)
    \put(200,204){$\times$}
    \bezier{2000}(50,0)(200,415)(350,0) 
    \put(93,103){$\bullet$}
    \put(172,164){$\times$}
    \bezier{1600}(93.5,103)(210,270)(300,0) 
    \put(202,126.5){$\times$}
    \bezier{1600}(100,0)(200,260)(330,0)  
    \put(280,10){\bf 1}
    \put(247.5,110){$\bullet$}
    \put(255,116){${}_{\ket{T}}$}
    \put(115.5,40){$\circ$}
    \put(123,40){${}_{S^-}$}
    \put(304,40){$\circ$}
    \put(293,40){${}_{S^+}$}
    \put(263,95){\Large$\ast$}
    \put(257,107.5){\vector(-1,1){4}}
    \put(195,130){${}^{\ket{s}}$}
    \put(73,104){${}^{\ket{E}_{\rm ch}}$}
    {\linethickness{1.5pt}
      \bezier{50}(264.5,99)(200,150)(120,0)
      }
    \put(220,95){${}^{\ket{\rm Sub}}$}
    \put(225,109){$\times$}
    \put(132,5){\bf 2}
  \end{picture}
  \label{onesided}
  \message{done}
\end{equation}
For brevity, we have denoted
$\ket{S}^\pm=\ket{S(r,1,\hcm(r,1,h,t),t)}^\pm$ and $\ket{E}_{\rm
  ch}=\ket{E(n,\hmc(r,1,h,t),t)}_{\rm ch}$.  Then,
$\ket{S}^-$---$\ket{S}^+$ is the extremal diagram of the massive Verma
submodule $\,\mU'$, the crosses denote top-level representatives,
$\bullet$s are (twisted) topological \hw{} states, and
$\ast=g(r-n+1,r-n-1)\,\ket{T}$ is the state in $\,\mU'$ from which the
action of $\cG_{r-n}$ produces the~$\ket{T}$ state.  The line
$\ket{T}$---$\ket{S}^+$ cannot be reached by the action of elements of
the $\N2$ algebra on~$\ket{T}$, since the arrow in the diagram that
represents the action of $\cG_{r-n}$ cannot be inverted because of the
topological \hw{} conditions satisfied by $\ket{T}$.  Instead, acting
with the highest of modes of $\cQ$ that produces a non-vanishing
result, one spans out the branch $\ket{T}$--{\bf 1}.

The subsingular vector emerges whenever it happens that the top-level
representative $\ket{s}$ does not generate all of the $\,\mU'$
submodule. Namely, assume that $r>n$ (this is actually the case in the
diagram drawn above). Then, neither $\ket{T}$ nor the top-level
representative $\ket{s}$ generate the states on the line
$\ket{T}$--$S^+$ and, therefore, the corresponding part of the massive
Verma submodule.  Restricting oneself to only the top-level
representatives of extremal diagrams as singular vectors, one is
limited, therefore, to the submodule generated from the top-level
vector $\ket{s}$. After taking the quotient with respect to the
singular vector $\ket{S(r,1,\hcm,t)}^-$ (or, {\it equivalently\/},
$\ket{s}$), one is left with the submodule whose extremal diagram is
precisely the line {\bf 2}--$\ket{S}^+$. The $\ast$ state becomes a
twisted topological \hw{} state --- the cusp in the dotted extremal
diagram. Therefore, $\ket{\rm Sub}$ becomes a {\it sub\/}singular
vector.  We see, however, that the appearance of a subsingular vector
is entirely due to choosing an inconvenient definition of singular
vectors: the problem with top-level representatives of (extremal
diagrams of) submodules is that they do not necessarily generate
maximal submodules.

\paragraph{Two singular vectors in the same grade.\label{sec:twoSV}}
As another instructive example of degeneration of massive Verma
modules, we consider the case where two linearly independent singular
vectors exist in the same grade. This effect was first observed
in~\cite{[Doerr2]}.  We consider here the simplest explicit example
(see~\cite{[ST4]} for the complete description).

In the module $\,\mU_{h, \ell, t}$, two linearly independent singular
vectors exist in the same grade when~\cite{[ST4]}
\begin{equation}
  h = \frac{(1 - m + n)s}{-m - n + r}\,,\quad
  t = \frac{-m - n + r}{s}\neq0\,,\qquad
  m,n\in\oN\,, \qquad
  2r-m-n\geq0
  \label{therules}
\end{equation}
Then,
\begin{equation}
  \theta_1(r,s,h,t)\bigm|_{\req{therules}} = -m\,,\qquad
  \theta_2(r,s,h,t)\bigm|_{\req{therules}} = n
\end{equation}
and the above formulas \req{Sgen} take the following form:
\begin{equation}\kern-8pt\new
  \begin{array}{rcl}
    \ket{S(r,s,h,t)}^-\Bigm|_{\req{therules}}\kern-6pt&=&\kern-6pt
    \cG_{-rs}\ldots\cG_{r-m-1}\,
    \cE^{-,-m}(r,s,t)\, \cG_{-m}\ldots\cG_{-1}\cdot
    \ket{h,\theel(r,s,h,t),t}\Bigm|_{\req{therules}}\,,
    \\
    \ket{S(r,s,h,t)}^+\Bigm|_{\req{therules}}\kern-6pt&=&\kern-6pt
    \cQ_{1-rs}\ldots\cQ_{r-n-1}\,
    \cE^{+,n}(r,s,t)\,\cQ_{-n}\ldots \cQ_{0}\cdot
    \ket{h,\theel(r,s,h,t),t}\Bigm|_{\req{therules}}\,,
  \end{array}
  \label{Sspecial}
\end{equation}

Obviously, now that the $g$ and $q$ operators from \req{Sgen} have
become the products of the $\N2$ generators, one observes that
$\ket{S(r,s,h,t)}^\pm$ are dense $\cG/\cQ$-descendants of the
respective vectors $\ket{T}^\pm$ given by
\begin{equation}\new
  \begin{array}{rcl}
    \ket{T}^-\kern-6pt&=&\kern-6pt \cE^{-,-m}(r,s,t)\,
    \cG_{-m}\ldots\cG_{-1}\,
    \ket{h,\theel(r,s,h,t),t}\Bigm|_{\req{therules}}\,,
    \\
    \ket{T}^+\kern-6pt&=&\kern-6pt \cE^{+,n}(r,s,t)\,
    \cQ_{-n}\ldots\cQ_{0}\,
    \ket{h,\theel(r,s,h,t),t}\Bigm|_{\req{therules}}\,,
  \end{array}
  \label{Sspecial2}
\end{equation}
and each of which satisfies the twisted {\it topological\/} \hw{}
conditions:
\begin{equation}\new
  \begin{array}{rclcrcl}
    \cG_{r-m}\ket{T}^-&=&0\,,&&\cQ_{m-r}\ket{T}^-&=&0\,,\\
    \cG_{n-r}\ket{T}^+&=&0\,,&&\cQ_{r-n}\ket{T}^+&=&0\,.
  \end{array}
\end{equation}

Let $\mC_-$ and $\mC_+$ be the submodules generated from the
respective vectors~\req{Sspecial2}. There exist $2r-m-n$ states
$\ket{x_i}^-$ from the extremal diagram of the $\mC_-$ submodule and
the same number of states $\ket{x_i}^+$ from the extremal diagram of
$\mC_+$ such that, for each $i$, $\ket{x_i}^-$ and $\ket{x_i}^+$ are
in the same grade. As all of the states on the extremal diagram of a
(twisted) topological Verma module, these states satisfy twisted
massive \hw{} conditions.

As a simple example, consider the case $r=2$, $s=1$, $m=2$, $n=1$.
There are two charged singular vectors in the corresponding massive
Verma module~$\,\mU_{0,-2,-1}$.  In each of the respective twisted
topological Verma modules, there exist topological singular vectors
directly on the extremal diagrams.  These two topological singular
vectors have identical $\cL_0$- and $\cH_0$-gradings (the dimension
and the $U(1)$ charge) and are linearly independent. They are shown by
the double bullet in the following diagram:
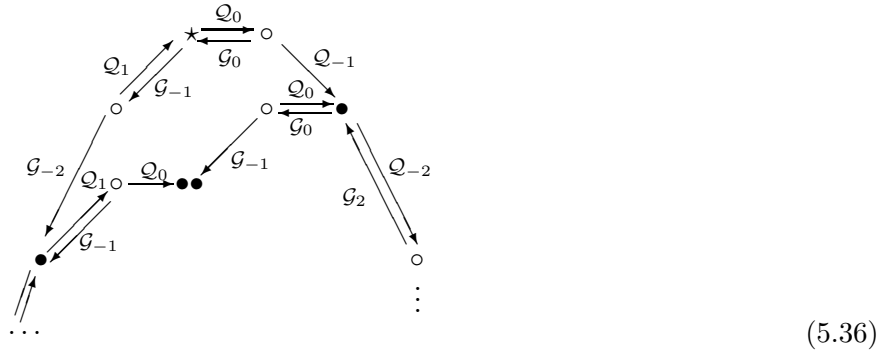
\begin{equation}
  \unitlength=1.00mm
  \begin{picture}(140,46)
    \put(50.00,10.00){
      \put(00.00,00.00){$\bullet$}
      \put(10.00,20.00){$\circ$}
      \put(20.00,30.00){$\star$}
      \put(30.00,30.00){$\circ$}
      \put(40.00,20.00){$\bullet$}
      \put(50.00,00.00){$\circ$}
      \put(09.50,20.00){\vector(-1,-2){8}}
      \put(11.50,23.00){\vector(1,1){7}}
      \put(19.70,29.00){\vector(-1,-1){7}}
      \put(22.20,31.50){\vector(1,0){7}}
      \put(28.70,29.95){\vector(-1,0){7}}
      \put(33.00,29.50){\vector(1,-1){7}}
      \put(43.00,19.00){\vector(1,-2){8}}
      \put(49.50,03.00){\vector(-1,2){8}}
      \put(-1.00,13.00){${}_{\cG_{-2}}$}
      \put(09.00,26.50){${}_{\cQ_{1}}$}
      \put(16.00,21.50){${}^{\cG_{-1}}$}
      \put(23.90,33.50){${}_{\cQ_{0}}$}
      \put(24.50,25.50){${}^{\cG_{0}}$}
      \put(37.00,27.50){${}_{\cQ_{-1}}$}
      \put(47.00,13.00){${}_{\cQ_{-2}}$}
      \put(41.00,07.00){${}^{\cG_{2}}$}
      \put(06.00,09.30){${}^{\cQ_1}$}
      \put(06.00,01.00){${}^{\cG_{-1}}$}
      \put(01.80,02.00){\vector(1,1){8}}
      \put(10.20,08.60){\vector(-1,-1){8}}
      \put(10.00,10.00){$\circ$}
      \put(12.60,11.00){\vector(1,0){6}}
      \put(18.80,10.00){$\bullet\bullet$}
      \put(26.20,12.00){${}^{\cG_{-1}}$}
      \put(39.50,20.40){\vector(-1,0){7}}
      \put(32.80,21.60){\vector(1,0){7}}
      \put(30.00,20.00){$\circ$}
      \put(29.70,19.70){\vector(-1,-1){7.5}}
      \put(33.50,23.40){${}_{\cQ_0}$}
      \put(34.00,18.40){${}_{\cG_0}$}
      \put(14.00,12.50){${}_{\cQ_0}$}
      \put(50.50,-06.00){$\vdots$}
      \put(-0.50,-0.50){\line(-1,-3){2}}
      \put(-1.80,-7.40){\vector(1,3){2}}
      \put(-3.50,-09.00){$\ldots$}
      }
  \end{picture}
  \label{degendiagram}
\end{equation}
(where the filled dots show twisted {\it topological\/} \hw{} states).

{}From this diagram we read off what the formulae \req{Sspecial2} give
us in this particular case:
\begin{equation}\new
  \begin{array}{rcl}
    \ket{T}^-&=&\cQ_0\,\cQ_1\,\cG_{-2}\,\cG_{-1}\,\ket{0,-2,-1}\,,\\
    \ket{T}^+&=&\cG_{-1}\,\cG_0\,Q_{-1}\,\cQ_0\,\ket{0,-2,-1}\,.
  \end{array}
  \label{example2}
\end{equation}
or, after some rearrangements,
\begin{equation}\kern-5pt\new
  \begin{array}{rcl}
    \ket{T}^-\kern-6pt&=&\kern-6pt\Bigl(8\cL_{-2} - 4\cG_{-2}\cQ_{0} -
    4\cH_{-1}  \cL_{-1}   +
    4\cL_{-1}  \cL_{-1}   -
    2\cQ_{-1}  \cG_{-1}   +
    2\cH_{-1}  \cG_{-1}  \cQ_{0}   -
    2\cL_{-1}  \cG_{-1}  \cQ_{0}\Bigr)\ket{0,-2,-1}  \,,\\
    \ket{T}^+\kern-6pt&=&\kern-6pt\Bigl(8 \cH_{-2} + 8 \cL_{-2} -
    4 \cQ_{-1}  \cG_{-1}   +
    2 \cG_{-1}  \cH_{-1}  \cQ_{0}   +
    2 \cG_{-1}  \cL_{-1}  \cQ_{0}\Bigr)\ket{0,-2,-1}\,.
  \end{array}
  \label{rexample2}
\end{equation}

\section{Conclusions and an outlook\label{sec:Concl}}\lvm
The construction of the $\N2$ singular vectors has been presented in a
way that makes it parallel to the known constructions for singular
vectors of the affine Lie algebras.  It is the {\it topological\/}
singular vectors of the $\N2$ algebra with central charge $\ctop\neq3$
that are in a 1:1 correspondence with singular vectors of the affine
$\SL2$ algebra of level $k$, $\ctop=3k/(k+2)$.

\smallskip

The analysis of the monomial expressions for the singular vectors
suggests that rather than identifying the three classes of singular
vectors (as in section \ref{sec:Prelim}) and then analyzing their
intersections, it may be useful to view the set of the $\N2$ singular
vectors as `stratified' according to which lengths of the continued
operators become integers in {\it either\/} of the formulae
\req{Sgen}, as we move from right to left. This approach is further
developed in~\cite{[ST4]}, with the embedding diagrams of $\N2$ Verma
modules classified and constructed in~\cite{[SSi]}.

\smallskip

It would be interesting to apply the present construction to the
various {\it realizations\/} of the $\N2$ algebra. In that respect, it
would be quite useful to have the conformal field-theoretic
counterparts of the operators \req{gandq}. The $\N2$ algebra will then
be represented by conformal fields $\cT(z)$, $\cH(z)$, $\cG(z)$, and
$\cQ(z)$.  Denoting the respective `continued' field operators as
$\sG_{\mu,\nu}(z)$ and $\sQ_{\mu,\nu}(z)$, with complex $\mu$ and
$\nu$, we would have the following counterpart of the `reduction'
rule~\req{integrallength}:
\begin{equation}
  \sG_{\lambda+N,\lambda}(z)=\cG^{(\lambda+N)}(z)\,\ldots\,\cG^{(\lambda)}(z)\
  {\rm for} \ N\in\oN_0\,,
\end{equation}
where $\cG^{(\mu)}(z)$ is the $\mu$-th derivative of $\cG(z)$; for
arbitrary complex $\mu$ it can be defined by analytically continuing
the integral
\begin{equation}
  f^{(N)}(z)={1\over N!}\,{1\over2\pi i}\oint{f(u)\,du\over(z-u)^{N+1}}
\end{equation}

The operator products of the bosonic fields $\cT$ and $\cH$ with $\sG$
are easily found, e.g.,
\begin{equation}\new
  \begin{array}{rcl}
    \cH(z)\sG_{\mu,\nu}(w)&=&(\mu-\nu+1)\,{\sG_{\mu,\nu}(w)\over z-w}\,,\\
    \cT(z)\sG_{\mu,0}(w)&=&\half(\mu+1)(\mu+4)\,{\sG_{\mu,0}(w)\over(z-w)^2}
    +{\d\sG_{\mu,0}(w)\over z-w}
  \end{array}
\end{equation}
Therefore $\sG_{\mu,0}$ are primary fields, an obvious translation of
the fact that the operators \req{gandq} map highest-weight states into
extremal states.  Further, we have the following version of the gluing
rules \req{Glue}:
\begin{equation}
  \sG_{\mu,\nu}(z)\,\sG_{\nu-1,\lambda}(z)=\sG_{\mu,\lambda}(z)\,.
\end{equation}

The positive powers in the short-distance expansion are controlled by
the following property of the $\sG$ operators:
\begin{equation}
  \d\sG_{\mu,\nu}(w)=\cG^{(\mu+1)}(w)\,\sG_{\mu-1,\nu}(w)\,,
  \label{derivative}
\end{equation}
which allows one to evaluate the higher-order derivatives recursively.
Thus, it is easily seen that
\begin{equation}
  \d^N\sG_{\mu,\nu}(w)=
  \d^N(\cG^{(\mu)}(w)\,\ldots\,\cG^{(\mu-N+1)}(w))\,\sG_{\mu-N,\nu}(w)\,.
\end{equation}

The fields $\sQ_{\mu,\nu}(z)$ satisfy the obvious analogues of the
above equations.

When considering various special realizations of the $\N2$ algebra,
the behaviour of $g(a,b)$ and $q(a,b)$ can be very complicated, while
evaluating the corresponding realizations of the field operators like
$\sG_{\mu,\nu}(z)$ can be much easier. One can thus consider the
Kazama--Suzuki mapping relating the affine $\SL2$ and $\N2$ algebras
and {\it extend\/} it to include the continued operators.  This will
then serve to find the pullbacks of the $\N2$ singular vectors in the
{\it monomial\/} forms, without having to rewrite them in the
conventional Verma forms. Thus, in particular, the continued
Kazama--Suzuki mapping relates the topological singular vectors in the
monomial form \req{Tplus}, \req{Tminus} to the respective MFF
monomials~\cite{[MFF]}. The construction for $\N2$ singular vectors
given in this paper is further be applied in~\cite{[FST]} to
constructing a functor between categories of \hw-type modules over
affine $\SL2$ and $\N2$ algebras.

\medskip

It is interesting whether the continued operators $g$ and $q$ can be
represented in the spirit of \cite{[FM]}, as integral operators acting
on an appropriate manifold.  Such a representation would then give an
explicit integral representation for the $\N2$ correlation functions.

\paragraph{Acknowledgements.}
We are particularly grateful to B.~Feigin and J.~Figueroa-O'Farrill
for very helpful discussions.  We would like to thank P.~Bowcock,
M.~Soloviev, A.~Taormina, K.~Thielemans, M.~Vasiliev, and B.~Voronov
for interesting discussions.  AMS is grateful to Ed Corrigan for the
kind hospitality at the Department of Mathematical Sciences,
University of Durham, where a part of this paper was written.  The
authors wish to thank the referee for correcting several inaccuracies.
The work of AMS was supported in part by grant \#93-0633 from the
European Community and the work of IYT, by the RFFI grant 96-01-00725.
The work of IYT is also supported by a Landau grant.

\end{document}